\documentclass[11pt]{article}
\usepackage{graphicx}

\def \bx{{\bf x}}

\def \bv{{\bf v}}

\def \bq{{\bf q}}

\def \Db{{\bf D}}
\def \Eb{{\bf E}}

\def \be{\begin{equation}} 
\def \ee{\end{equation}} 
\def \d{\partial} 
\def \aa{\alpha} 
\def \bb{\beta} 
\def \dd{{\delta}} 
\def \gg{\gamma} 
\def \ss{\sigma} 
\def \ll{\lambda} 
\def \om{\omega}
\def \th{\theta}

\def \th{\theta}
\def \eps{\epsilon}
\def \kk{\kappa}
\def \DD{\Delta}
\def \SS{\Sigma} 
\def \LL{\Lambda}
\def \GG{\Gamma}
\def \Om{\Omega}
\def \Th{\Theta}
\def \Thb{{\bar \Th}}

\def \Ec{{\cal E}}
\def \Mc{{\cal M}}

\def \Lc{{\cal L}}
\def \la{\langle} 
\def \ra{\rangle} 
\def \fr{\frac}

\def \ab{{\bar a}}
\def \db{{\bar d}}
\def \fb{{\bar f}}
\def \hb{{\bar h}}
\def \pb{{\bar p}}

\def \yb{{\bar y}}
\def \fb{{\bar f}}
\def \gb{{\bar g}}
\def \ggb{{\bar \gg}}
\def \ght{{\hat g}}
\def \gght{{\hat \gg}}
\def \eth{{\hat \eta}}
\def \eb{{\bar e}}

\def \mb{{\bar m}}
\def \mub{{\bar \mu}}
\def \llb{{\bar \ll}}
\def \thb{{\bar \theta}}
\def \chib{{\bar \chi}}
\def \Abr{{\bar A}}
\def \Db{{\bar D}}
\def \Eb{{\bar E}}
\def \Lb{{\bar L}}
\def \Gb{{\bar G}}
\def \Mb{{\bar M}}
\def \Nb{{\bar N}}
\def \Omb{{\bar \Omega}}
\def \SSb{{\bar \SS}}
\def \LLb{{\bar \Lambda}}
 
\def \st{{\star}}

\begin{document} 
\title{Quantum Field Theory in a Multi-Metric Background } 
\author{I.T. Drummond\thanks{email: itd@damtp.cam.ac.uk} \\ 
            Department of Applied Mathematics and Theoretical Physics\\ 
            Centre for Mathematical Sciences\\ 
            Wilberforce Road\\ Cambridge\\ England, CB3 0WA
        }
\maketitle
\abstract{
By means of simple models in a flat spacetime manifold we examine some of the issues that
arise when quantizing interacting quantum fields in multi-metric backgrounds.
In particular we investigate the maintenance of a causal structure in the models.
In this context we introduce and explain the relevance of an interpolating metric 
that is a superposition of the individual metrics in the models. We study the 
renormalisation of a model with quartic interactions and elucidate the structure 
of the renormalisation group and its implications for Lorentz symmetry breakdown.
}

\vfill
DAMTP-2013-13
\pagebreak

\section{\bf Introduction}

In standard General Relativity the structure of the space-time manifold is expressed
in terms of a metric field. This metric provides a lightcone structure that determines causal 
relationships between events in the manifold. Modified gravity theories involving more than 
one metric have been proposed as potential explanations of dark matter and dark energy effects 
in cosmology \cite{MOFF,MAG,MILG,ITD}. They are also relevant to models of massive gravity\cite{HINT,HASS}. 
A comprehensive report on work in this area can be found in reference \cite{FER}. 
While these approaches are for the moment conjectural they are of considerable interest in themselves.

An important feature of such theories is that they may lead to a violation
of Lorentz invariance. There have been many careful studies of the breakdown of Lorentz 
invariance \cite{COLGL1, KOST1, KOST2} with the conclusion that observations constrain it 
to be a very small effect indeed \cite{KOST3}. Nevertheless it is interesting to consider 
what issues of principle are involved in dealing with quantum field theory in the context 
of a multi-metric space-time without any {\it a priori} constraint on the relationship of 
the various metrics except for the general requirement of maintaining a causal structure 
for the theory. These issues of causality and multiple lightcone structure appear in
in other contexts \cite{ITDSJH}, \cite{SHHO1,SHHO2,SHHO3}, \cite{GibHer}.

In this paper we are concerned not with the gravitational dynamics implied by
such theories but with the behaviour of quantum fields in the presence of
more than one space-time metric. We therefore simplify the problem and consider
a {\it flat} space-time with several metrics each of which is associated with a
particular quantum field. We investigate the constraints resulting from a requirement
of causality by studying some simple scalar field theory models. In the simplest case 
we have a model with two fields and two metrics. We
find that if we wish to enforce causality when there are interactions between the
fields, it is important that the two metrics support lightcones that overlap and 
share spacetime vectors that are timelike in both metrics. A useful and significant construct 
that plays a role in elucidating the relationship between the two metrics is that of the
{\it interpolating metric} which is built from a linear superposition with positive weights 
of the two metrics. As the weights of the superposition vary the interpolating metric 
passes from one metric to the other and may at an intermediate stage become singular. 
The avoidance of this singularity is important for the dynamics of the theory and leads 
to the restriction indicated above requiring the two lightcones to intersect by sharing a 
set of spacetime vectors that are timelike in both metrics. We indicate briefly generalisations 
to cases with more than two metrics.

For a model in which the scalar fields experience quartic interactions we investigate
the renormalisation of the perturbation series. The interpolating metric plays a
crucial role in the evaluation of the of the Feynman diagrams of the theory reinforcing
the restrictions on the lightcones required for a causal structure. The
individual metrics are renormalised as part of the calculation and are involved
through the properties of the interpolating metric in the renormalisation group flows
describing the evolution of coupling constants and masses in the theory.

\section{\label{MMT} Multi-Metric Theory}

The fundamental assumption we make is that there exists a flat space-time manifold
the events in which are specified by a set of coordinates $x^\mu$, $\{\mu=0,1,2,3\}$
that are arbitrary under linear transformations. On the manifold are a set of $N$
metrics $g^{(i)}_{\mu\nu}$, $\{i=1,2,\cdots,N\}$ that are independent
of $x^\mu$~. Each metric specifies a lightcone on the manifold through the
equation
\be
g^{(i)}_{\mu\nu}x^\mu x^\nu=0~~.
\label{lightcone}
\ee
That eq(\ref{lightcone}) does determine a lightcone is ensured by constructing
$g^{(i)}_{\mu\nu}$ from an associated vierbein $e^{(i)}_{a\mu}$ by means of the standard
Minkowski metric $\eta^{ab}$ with diagonal entries $\{1,-1,-1,-1\}$ thus
\be
g^{(i)}_{\mu\nu}=\eta_{ab}e^{(i)a}_{~~~~\mu}e^{(i)b}_{~~~~\nu}~~.
\ee
Of course the vierbein is arbitrary under Lorentz transformations, $L^a_{~~b}$,
\be
e^{(i)a}_{~~~~\mu}\rightarrow L^a_{~~b}e^{(i)a}_{~~~~\mu}~~.
\ee
With each vierbein we associate a set of coordinates $y^{(i)a}=e^{(i)a}_{~~~~\mu}x^\mu$~.
We also have the inverse relation $x^\mu=e^{(i)\mu}_{~~~~a}y^{(i)a}$~.
The space-time volume elements are related by
\be
d^4y^{(i)}=\Om^{(i)}d^4x~~,
\ee
where
\be
\Om^{(i)}=\det e^{(i)a}_{~~~~\mu}~~.
\ee
A plane wave in space-time has the form
\be
f_q(x)=e^{-iq_\mu x^\mu}~~.
\ee
Viewed from the vierbein frame $(i)$ the four vector
is 
\be
p^{(i)a}=e^{(i)\mu a}q_\mu~~.
\ee
Obviously
\be
q_\mu x^\mu=p^{(i)}_{~~~a}y^{(i)a}~~,
\ee
and
\be
g^{(i)\mu\nu}q_\mu q_\nu=\eta_{ab}p^{(i)a}p^{(i)b}~~.
\ee

Our aim is to investigate models of quantum field theory in which each lightcone
is associated with a set of fields that transform in a conventional fashion under 
a Lorentz group associated with that lightcone but which may also interact with
fields associated with other lightcones. This is a scalar field version of a 
model investigated by Colemam and Glashow \cite{COLGL2} in relation to neutrino mixing.
There are then two issues. The first is that 
there is {\it in general} no overall Lorentz invariance for such a theory. There may however
be circumstances, as suggested by Coleman and Glashow \cite{COLGL1}, in which a reduced Lorentz invariance
survives.

\begin{figure}[t]
   \centering
  \includegraphics[width=0.6\linewidth]{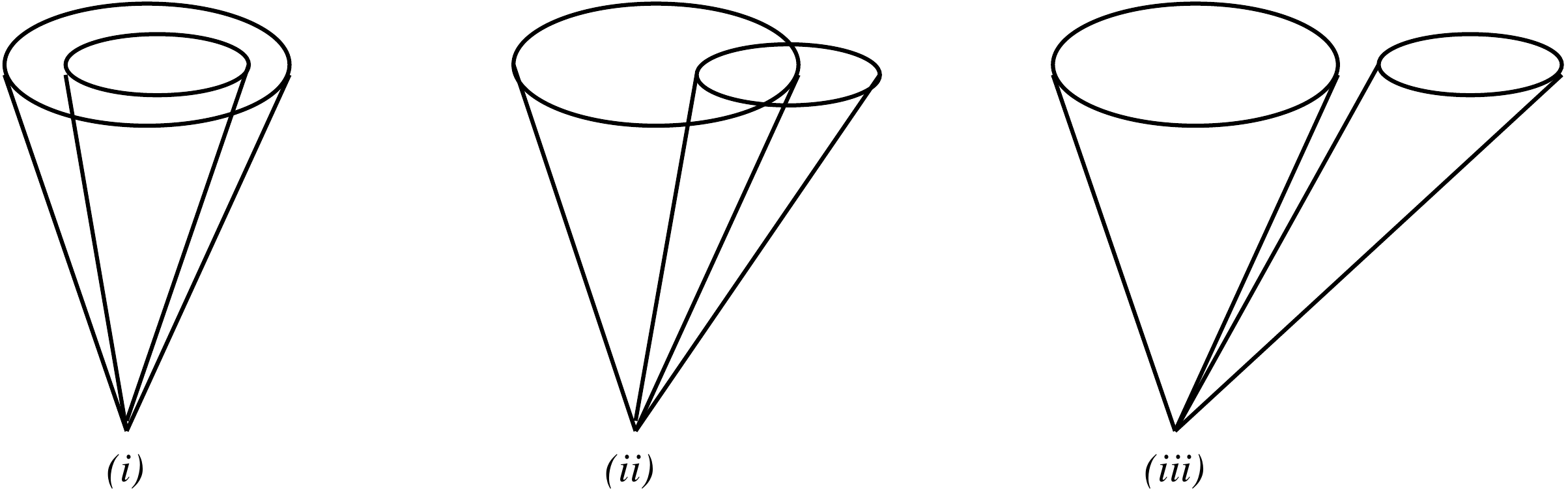}
  \caption{Examples of pairs of lightcones (in 2+1 dimensions)
that share spacelike surfaces. The first two examples represent situations in which there
are spacetime vectors that are timelike in both cones. In the third case each lightcone is
seen as spacelike from the other. }
  \label{LCONES}
\end{figure}

The second issue is that we require restrictions on the relationship between the lightcones in order
that there can be a causal evolution of the fields. A natural way of achieving this is to require
that there exists a foliation of the (flat) space-time manifold that is space-like
with respect to each of the lightcones simultaneously. An example of the possible relationships
between two lightcones that respect this constraint are illustrared in Fig \ref{LCONES}. 

We will identify  
the coordinate that labels the leaves of the foliation, as the time variable. Of course if there is
one such foliation we expect there will normally be an infinite set. This will become clearer
in the example of two lightcones described in section \ref{BMG}. 
When we come to consider the dynamics of interacting quantum field theory models in more detail we will
be led to impose further restrictions on the relationship between the lightcones.

In order to emphasize the importance of the space-time manifold we will formulate 
the theory so that it is invariant under general linear transformations of the 
coordinates on the manifold.

\section{\label{BMG}Bi-Metric Geometry}

The basic ideas can be understood by considering a model in which there are two metrics.
We simplify the notation and introduce metrics $g_{\mu\nu}$ and $\gb_{\mu\nu}$ that are  
constructed from two vierbeins $e_{a\mu}$ and $\eb_{a\mu}$~.
\begin{eqnarray}
g_{\mu\nu}&=&\eta_{ab}e^a_{~~\mu}e^b_{~~\nu}~~,\nonumber\\
\gb_{\mu\nu}&=&\eta_{ab}\eb^a_{~~\mu}\eb^b_{~~\nu}~~.
\end{eqnarray}

This guarantees that the two surfaces
\begin{eqnarray}
g_{\mu\nu}x^\mu x^\nu&=&0~~,\nonumber\\
\gb_{\mu\nu}x^\mu x^\nu&=&0~~,
\end{eqnarray}
are indeed lightcones of conventional form. The relationship between the two lightcones
determines the nature of the breakdown of Lorentz invariance in the theory.
We have
\begin{eqnarray}
\Om&=&\det e^a_{~~\mu}~~,\nonumber\\
\Omb&=&\det \eb^a_{~~\mu}~~.
\end{eqnarray}
The inverse vierbeins are $e^\mu_{~~ a}$ and $\eb^\mu_{~~ a}$ and satisfy
\begin{eqnarray}
e^a_{~~\mu}e^\mu_{~~b}&=&\dd^a_b~~,\nonumber\\
\eb^a_{~~\mu}\eb^\mu_{~~b}&=&\dd^a_b~~.
\end{eqnarray} 
We can construct coordinates appropriate to each vierbein,
\begin{eqnarray}
y^a&=&e^a_{~~\mu}x^\mu~~,\nonumber\\
\yb^a&=&\eb^a_{~~\mu}x^\mu~~.
\end{eqnarray}
The inverse transformations are
\begin{eqnarray}
x^\mu&=&e^\mu_{~~a}y^a~~,\nonumber\\
x^\mu&=&\eb^\mu_{~~a}\yb^a~~.
\end{eqnarray}
Under a Lorentz transformation $L^a_{~~b}$ 
\be
y^a\rightarrow L^a_{~~b}y^b~~.
\ee
Now
\be
e^a_{~~\mu}\rightarrow L^a_{~~b}e^b_{~~\mu}=e^a_{~~\nu}\LL^\nu_{~~\mu}~~,
\ee
where therefore $\LL^\nu_{~~\mu}$ is also a representation of the Lorentz group
that preserves the metric $g_{\mu\nu}$~. Under the same Lorentz transformation
it follows that
\be
x^\mu\rightarrow \LL^\mu_{~~\nu} x^\nu~~.
\ee
Similarly for the vierbein $\eb_{a\mu}$, we have associated coordinates $\yb^a$ where
\be
\yb^a=\eb^a_{~~\mu}x^\mu~~,
\ee
and under a Lorentz transformation $\Lb^a_{~~b}$
\be
\yb^a\rightarrow \Lb^a_{~~b}\yb^b~~,
\ee
with
\be
x^\mu\rightarrow \LLb^\mu_{~~\nu} x^\nu~~,
\ee
where
\be
\eb^a_{~~\mu}\rightarrow \Lb^a_{~~b}\eb^b_{~~\mu}=\eb^a_{~~\nu}\LLb^\nu_{~~\mu}~~.
\ee
It follows that $\LLb^\nu_{~~\mu}$ is a representation of the Lorentz group
that preserves the metric $\gb_{\mu\nu}$~. It is obvious that in general 
$\LL^\mu_{~~\nu}$ and $\LLb^\mu_{~~\nu}$ are not the same representations of the Lorentz 
group since they preserve different metrics on space-time. As indicated in section \ref{MMT}
there can be circumstances in which the two versions of the Lorentz group
share a subgroup \cite{COLGL1,COLGL2}. This subgroup will then provide a reduced Lorentz invariance of the overall theory.

\subsection{\label{FLT} Foliation and Linear Transformations}

The coordinate space of the flat manifold is identical with the tangent space (at the origin).
In order to set up a foliation of the coordinate vector space we choose a cotangent vector $n_\mu$
and define a subspace $X$ by the restriction
\be
n_\mu x^\mu=0~~.
\ee
We then introduce a vector $m^\mu$ not lying in $X$ which satisfies
\be
n_\mu m^\mu=1~~.
\ee
A general vector $x^\mu$ can be expressed in the form
\be
x^\mu=\tau m^\mu+y^\mu~~,
\ee
where $y^\mu$ lies in $X$~. It follows that
\be
n_\mu x^\mu=\tau~~.
\ee
The subsets of coordinate space each labelled by a value of $\tau$, constitute a foliation.

It is useful to parametrize linear transformations on coordinate space in a manner
consistent with the foliation. First we introduce general linear transformation $M^\mu_{~~\nu}$
on the subspace $X$ which satisfies
\be
n_\mu M^\mu_{~~\nu}=n_\nu~~,
\ee
and
\be
M^\mu_{~~\nu} y^\nu=y^\mu~~.
\ee
This transformation maps $X$ into itself and leaves $\tau$ unchanged.
We follow with a shear transformation $S(v)^\mu_{~~\nu}$ that has the form
\be
S^\mu_{~~\nu}=\dd^\mu_\nu-v^\mu n_\nu~~,
\ee
where $v^\mu$ is a vector lying in $X$, that is $n_\mu v^\mu=0$~. It follows that
\be
S^\mu_{~~\nu}(v)x^\nu=x^\mu-\tau v^\mu~~.
\ee
Clearly $S^\mu_{~~\nu}(v)$ leaves $\tau$ unaltered but shifts the origin of
coordinates lying in the subspace $X$ by an amount proportional to $\tau$~. We have also
\be
(S^{-1}(v))^\mu_{~~\nu}=S^\mu_{~~\nu}(-v)~~.
\ee

If we assume coordinate space has $n$ dimensions then $X$ has $n-1$ dimensions and the 
linear transformation $M^\mu_{~~\nu}$ has $(n-1)^2$ parameters. The shear transformation has 
$n-1$ parameters so the combined transformation $(S(v)M)^\mu_{~~\nu}$ has $n^2-n$
parameters. If we now apply an overall scale transformation, $\om\dd^\mu_\nu$
this introduces one further parameter, bringing the total to $n^2-n+1$~. This
procedure is the most general transformation that preserves the  the foliation 
determined by $n_\mu$ and $m^\mu$~. 
The remaining $n-1$ parameters needed to complete the parametrization of the 
a general transformation may be chosen to be the velocities of the $n-1$ appropriately 
constructed Lorentz boosts along $n-1$ directions within the the foliation subspace $X$. 

We emphasise that although the resulting parametrization of the linear transformation on coordinate space
is perfectly general for transformations in a neighbourhood of the identity it will not cover
the whole general linear group. It is however the appropriate parametrization for our purposes.

\subsection{\label{RbM} Relationship between Metrics}

The relationship between two metrics can be associated with a general linear
transformation of the coordinate vector $x^\mu$,
\be
x^\mu\rightarrow x'^\mu=A^\mu_{~~\nu}x^\nu~~,
\ee
and
\be
g_{\mu\nu}\rightarrow \gb_{\mu\nu}=A^\ss_{~~\mu}A^\tau_{~~\nu}g_{\ss\tau}~~.
\ee
The construction of metrics from vierbeins is an example of this procedure.

We can now see that it is convenient to parametrize $A^\mu_{~~\nu}$ in the following way,
\be
A^\mu_{~~\nu}=\om \LL^\mu_{~~\ss}S^\ss_{~~\ll}M^\ll_{~~\nu}~~,
\ee
where $\LL^\mu_{~~\nu}$ is a general Lorentz boost in a representation that leaves $g_{\mu\nu}$
invariant. This contains the remaining $n-1$ parameters needed to complete the total of $n^2$
for a $n$ dimensional general linear transformation, that is connected to the identity.
The advantage of this completion is that it leaves the relationship between the two metrics
unchanged. If we have identified $X$ as the $(n-1)$ dimensional subspace of displacements
that are spacelike with respect to both metrics then the final Lorentz boosts may be omitted
without losing generality.

Finally we note as indicated above, that the general vierbein (or $n$-bein) can be constructed
by the same sequence of steps, the starting metric being the standard Lorentz metric in this case.

\subsection{\label{INTPL} Interpolating Metrics}

A further construct that is important in reflecting aspects of the relationship between two metrics
is that of the {\it interpolating metric}. 
It turns out to play a crucial role in elucidating the causal structure of interacting quantum field theory
with two or more lightcones. 
We define two versions. The first is ${\hat g}^{\mu\nu}(u)$ where $0\le u\le 1$ and
\be
{\hat g}^{\mu\nu}(u)=ug^{\mu\nu}+(1-u)\gb^{\mu\nu}~~,
\label{intmet1}
\ee
with its inverse ${\hat g}_{\mu\nu}(u)$~. The second is ${\tilde g}_{\mu\nu}(u)$ where
\be
{\tilde g}_{\mu\nu}(u)=(1-u)g_{\mu\nu}+u\gb_{\mu\nu}~~,
\label{intmet2}
\ee
with its inverse ${\tilde g}^{\mu\nu}(u)$~.
These two interpolating metrics are not inverses of one another but do satisfy the relation
\be
{\hat g}^{\mu\nu}(u)=g^{\mu\ss}{\tilde g}_{\ss\tau}(u)\gb^{\tau\nu}~~.
\ee
It follows that 
\be
\det {\hat g}^{\mu\nu}(u)=\det g^{\mu\ss}\det {\hat g}_{\ss\tau}(u)\det \gb^{\tau\nu}~~,
\ee
and therefore that if one of these interpolating metrics becomes singular so does the other.
It turns out that ${\hat g}^{\mu\nu}(u)$ is more directly related to the dynamics of the field
theory but the behaviour of ${\tilde g}_{\ss\tau}(u)$ is more easily interpreted in terms of the 
the relationship between the coordinate lightcones.

The outcome of a detailed investigation is that if the two lightcones overlap so that there
are coordinate vectors that timelike with respect to both lightcones then the interpolating metrics 
remain nonsingular for $0<u<1$~. When no such vectors exist then the interpolating metrics do become 
singular for $u$ in this range. The transitional situation arises when the lightcones of the
original metrics touch one another along a shared light ray. We illustrate these circumstances
with a simple example in section \ref{SpecBi}. A more detailed analysis is presented in appendix \ref{SIM}.

\section{\label{SQFT} Simple Quantum Field Theory Model}

We construct a simple quantum field theory involving a real scalar field $\phi(x)$
that we associate with the vierbein $e^a_{~~\mu}$~. It has a Lagrangian density
\be
\Lc_\phi(x)=\fr{1}{2}\left(g^{\mu\nu}\d_\mu\phi(x)\d_\nu\phi(x)-m^2\phi^2(x)\right)~~.
\ee
The action is $S_\phi$ given by 
\be
S_\phi=\int d^4x\Om\Lc_\phi(x)~~.
\ee
Clearly at this stage the action is invariant under both Lorentz transformations of
the vierbein $e^a_{~~\mu}$ and general linear coordinate transformations. In the notation of
subsection \ref{FLT} we choose $X$ to be a spacelike subspace of the metric $g_{\mu\nu}$
and set $n_\mu=(1,0,0,0,)$ and $m^\mu=(1,0,0,0,)$~. The coordinate $x^0$ is the time
variable labelling the leaves of the foliation and the coordinates $\bx=(x^1,x^2,x^3)$
are coordinates on the spacelike leaves.

The quantization is achieved by introducing the field, $\pi(x)$, conjugate to
the field $\phi(x)$ in the standard way
\be
\pi(x)=\fr{\dd S_\phi}{\dd({\d_0\phi}(x))}
      =\Om g^{0\mu}\d_\mu\phi(x)=\Om\left(g^{00}\d_0\phi(x)+g^{0i}\d_i\phi(x)\right)~~.
\label{conj1}
\ee
We complete the procedure by imposing the equal time commutation relations 
\begin{eqnarray}
\left[\phi(x^0,\bx),\pi(x^0,\bx')\right]&=&i\dd(\bx-\bx')~~, \nonumber\\
\left[\phi(x^0,\bx),\phi(x^0,\bx')\right]&=&0~~, \nonumber\\
\left[\pi(x^0,\bx),\pi(x^0,\bx')\right]&=&0~~.
\label{comrel1}
\end{eqnarray}
The Hamiltonian generating time development is
\be
H_\phi=\int d^3\bx\left\{\pi(x^0,\bx)(\d_0\phi(x^0,\bx))-\Om\Lc_\phi(x^0,\bx)\right\}~~.
\ee
That is
\be
H_\phi=\int d^3\bx\left\{\fr{1}{2}\fr{1}{g^{00}\Om}\left(\pi(x)-\Om(g^{0i}\d_i\phi(x))\right)^2
               -\fr{1}{2}\Om g^{ij}\d_i\phi(x)\d_j\phi(x)+\fr{1}{2}m^2\Om\phi^2(x)\right\}~~.
\ee
It is straightforward to check that this Hamiltonian yields a time development for the system
that is the standard equation of motion derived directly from the Euler-Lagrange equations,
namely
\be
g^{\mu\nu}\d_\mu\d_\nu\phi(x)+m^2\phi^2(x)=0~~.
\label{eqmot1}
\ee
We can obtain solutions for $\phi(x)$ as superpositions
of plane-waves 
\be
\phi(x)=f_q(x)=e^{-iq_\mu x^\mu}~~,
\ee
where $q$ satisfies the mass-shell condition
\be
g^{\mu\nu}q_\mu q_\nu=m^2~~.
\label{shell1}
\ee
It is implicit that we select the branch of the mass-shell condition corresponding to 
positive energy in the appropriate vierbein frame. 
Note that because of our definitions the velocity of a particle wave-packet 
is $\bv=-\d q_0/\d \bq$~.

If $f(x)$ and $f'(x)$ both satisfy eq(\ref{eqmot1}) then the current
\be
J_\phi^\mu(x)=i\Om g^{\mu\nu}\left(f'^\st(x)\d_\nu f(x)-f(x)\d_\nu f'^\st(x)\right)~~,
\label{current1}
\ee
satisfies
\be
\d_\mu J_\phi^\mu(x)=0~~. 
\ee
This permits us to define the time-independent scalar product of two wave functions $f(x)$ and $f'(x)$ 
to be
\be
(f',f)=i\Om\int d^3\bx \left\{f'^\st(x) g^{0\nu}\d_\nu f(x)-f(x)g^{0\nu}\d_\nu f'^\st(x) \right\}~~,
\ee
We have then
\be
(f_{q'},f_q)=2\Om g^{0\nu}q_\nu(2\pi)^3\dd_(\bq-\bq')~~,
\ee
where $\bq$ is the spatial part of $q_\mu$~~.

As a superposition of plane waves the field $\phi(x)$ can be put in the form
\be
\phi(x)=\int \fr{d^3\bq}{2\Om g^{0\nu}q_\nu(2\pi)^3}( f_q(x)a(q)+f^\st_q(x)a^\dag(q))~~.
\ee
The mode operators $a(q)$ are given by
\begin{eqnarray}
a(q)&=&(f_q,\phi)\nonumber\\
     &=&i\Om\int d^3\bx(f_q^\st(x)g^{0\nu}\d_\nu\phi(x)-\phi(x)g^{0\nu}\d_\nu f^\st_q(x))\nonumber\\
     &=&i\int d^3\bx f_q^\st(x)(\pi(x)-i\Om g^{0\nu}q_\nu\phi(x))~~.
\end{eqnarray}
Similarly we have
\be
a^\dag(q) =-i\int d^3\bx f_q(x)(\pi(x)+i\Om g^{0\nu}q_\nu\phi(x))~~.
\ee
The commutation relations in eq(\ref{comrel1}) imply
\begin{eqnarray}
\left[a(q),a(q')\right]&=&0~~,\nonumber\\
\left[a^\dag(q),a^\dag(q')\right]&=&0~~,\nonumber\\
\left[a(q),a^\dag(q')\right]&=&2\Om g^{0\nu}q_\nu(2\pi)^3\dd(\bq-\bq')~~.
\label{comms1}
\end{eqnarray}

The Hamiltonian can be expressed in terms of the mode operators (after normal ordering)
in the form
\be
H_\phi=\int \fr{d^3\bq}{(2\pi)^3\Om2g^{0\nu}q_\nu}q_0a^\dag(q)a(q)~~.
\label{HAM1}
\ee 

The Feynman propagator is
\be
G_F(x-x')=-i\la 0|T(\phi(x)\phi(x'))|0\ra~~.
\label{Feynman1}
\ee
It satisfies the equation
\be
(g^{\mu\nu}\d_\mu\d_\nu+m^2)G_F(x-x')=-\fr{1}{\Om}\dd(x-x')~~,\ee
and hence can be put in the form
\be
G_F(x-x')=\fr{1}{\Om}\int\fr{d^4q}{(2\pi)^4}\fr{e^{-iq_\mu x^\mu}}{g^{\mu\nu}q_\mu q_\nu-m^2+i\eps}~~.
\ee
This analysis is no more than standard elementary quantum field theory (QFT-101!) written out
for general (linear) coordinates. 

We now introduce a second real field $\psi(x)$ associated with the vierbein $\eb_{a\mu}$
together with a Lagrangian density
\be
\Lc_\psi(x)=\fr{1}{2}\left\{\gb^{\mu\nu}\d_\mu\psi(x)\d_\nu\psi(x)-\mb^2\psi^2(x)\right\}~~.
\ee
The associated action is
\be
S_\psi=\int d^4x\Lc_\psi(x)~~.
\ee
We can carry out an analysis entirely parallel to that for the field $\phi(x)$~.
The field conjugate to $\psi(x)$ is $\om(x)$ where
\be
\om(x)=\fr{\dd S_\psi}{\dd({\d_0\psi}(x))}
      =\Omb \gb^{0\mu}\d_\mu\psi(x)=\Om\left(\gb^{00}\d_0\psi(x)+\gb^{0i}\d_i\psi(x)\right)~~.
\label{conj2}
\ee
The standard equal time commutation relations are imposed on $\psi(x)$ and $\om(x)$~.
The equation of motion for $\psi(x)$ is
\be
\gb^{\mu\nu}\d_\mu \d_\nu\psi(x)+\mb^2\psi(x)=0~~,
\label{eqmot2}
\ee
and the mode decomposition for $\psi(x)$ in the form
\be
\psi(x)=\int \fr{d^3\bq}{2\Om g^{0\nu}q_\nu(2\pi)^3}( \fb_q(x)b(q)+\fb^\st_q(x)b^\dag(q))~~.
\ee
The wavefunctions $\fb_q(x)$ are
\be
\fb_q(x)=e^{-iq_\mu x^\mu}~~,
\ee
where now $q$ satisfies the mass-shell condition
\be
\gb^{\mu\nu}q_\mu q_\nu=\mb^2~~.
\label{shell2}
\ee
The appropriate definition of the scalar product between wave functions has the form
\be
(\fb',\fb)=\int d^3\bx J_\psi^0(x)~~,
\ee
where the conserved current $J_\psi^\mu(x)$ is
\be
J_\psi^\mu(x)=i\Omb\gb^{\mu\nu}\left(\fb'^\st(x)\d_\nu \fb(x)-\fb(x)\d_\nu \fb'^\st(x)\right)~~.
\label{current2}
\ee

The Hamiltonian for the new field can be constructed in the form
\be
H_\psi=\int \fr{d^3\bq}{(2\pi)^3\Om 2\gb^{0\nu}q_\nu}q_0b^\dag(q)b(q)~~,
\label{HAM2}
\ee
where the $b$-coefficients satisfy the commutation relations
\begin{eqnarray}
\left[b(q),b(q')\right]&=&0~~,\nonumber\\
\left[b^\dag(q),b^\dag(q')\right]&=&0~~,\nonumber\\
\left[b(q),b^\dag(q')\right]&=&2\Omb \gb^{0\nu}q_\nu(2\pi)^3\dd(\bq-\bq')~~.
\label{comms2}
\end{eqnarray}

The Feynman propagator for $\psi(x)$ is
\be
\Gb_F(x-x')=\fr{1}{\Omb}\int\fr{d^4q}{(2\pi)^4}\fr{e^{-iq_\mu x^\mu}}{\gb^{\mu\nu}q_\mu q_\nu-\mb^2+i\eps}~~.
\label{Feynman2}
\ee

Assuming that the $a$-operators and $b$-operators commute we can construct a joint
system and obtain the action
\be
S=S_\phi+S_\psi~~.,
\ee
and the Hamiltonian
\be
H=H_\phi+H_\psi~~.
\label{HAM3}
\ee
Even though the two theories are individually Lorentz invariant the joint 
system does not in general, as indicated in section \ref{MMT}, exhibit Lorentz invariance.

\subsection{\label{MIX}Mixing Interaction}

In this non-Lorentz invariant context we introduce a mixing interaction between
the field $\phi(x)$ and $\psi(x)$ by adding further term to the action so it becomes
\be
S=S_\phi+S_\psi-w^2\int d^4x(\Om\Omb)^{\fr{1}{2}}\phi(x)\psi(x)~~.
\ee
We have included the factor $(\Om\Omb)^\fr{1}{2}$ with the integration weight in order to
maintain a formal symmetry between the two fields.

An immediate consequence of including the interaction between the fields is that the theory
is explicitly non-Lorentz invariant whenever the two lightcones do not coincide. 
We can still carry out the canonical quantization procedure for the joint system.
The fields $\pi(x)$ and $\om(x)$ conjugate to $\phi(x)$ and $\psi(x)$ respectively
are still given by eq(\ref{conj1}) and eq(\ref{conj2}) and we can still impose 
the standard commutation relations.

The equations of motion however are now
\begin{eqnarray}
\Om g^{\mu\nu}\d_\mu\d_\nu\phi(x)+\Om m^2\phi(x)+w^2(\Om\Omb)^{\fr{1}{2}}\psi(x)&=&0~~,\nonumber\\
\Omb\gb^{\mu\nu}\d_\mu\d_\nu\psi(x)+\Om \mb^2\psi(x)+w^2(\Om\Omb)^{\fr{1}{2}}\phi(x)&=&0~~.
\label{eqmot3}
\end{eqnarray}
Plane wave solutions are of the form
\be
F_q(x)=
\left(\begin{array}{c}
    \phi(x)\\
    \psi(x)
\end{array}\right)=\left(\begin{array}{c}
    f_q(x)\\
    \fb_q(x)
\end{array}\right)=\left(\begin{array}{c}
       A_q \\
       \Abr_q
\end{array}\right)e^{-iq_\mu x^\mu}~~.
\label{planemix}
\ee
The amplitudes $A_q$ and $\Abr_q$ satisfy
\begin{eqnarray}
\Om \left(g^{\mu\nu}q_\mu q_\nu -m^2\right)A_q-w^2(\Om\Omb)^{\fr{1}{2}}\Abr_q&=&0~~,\nonumber\\
\Omb\left(\gb^{\mu\nu}q_\mu q_\nu-\mb^2\right)\Abr_q-w^2(\Om\Omb)^{\fr{1}{2}}A_q&=&0~~.
\label{eqmot3a}
\end{eqnarray}
In order that these equations be soluble we must impose the determinantal condition
\be
\left(g^{\mu\nu}q_\mu q_\nu -m^2\right)\left(\gb^{\mu\nu}q_\mu q_\nu-\mb^2\right)-w^4=0~~.
\label{shell3}
\ee
This yields the generalization of the mass-shell conditions appropriate to the model.
When the mixing term is absent, $w^2=0$, this reduces to the two mass-shell conditions
in eq(\ref{shell1}) and eq(\ref{shell2})~. When $w^2\ne 0$ and typical components of $q_\mu$ 
are large then eq(\ref{shell3}) implies that either $(g^{\mu\nu}q_\mu q_\nu-m^2)$
or $(\gb^{\mu\nu}q_\mu q_\nu-\mb^2)$ is large but not both. If we assume that the second
of these is large then on writing eq(\ref{shell3}) in the form
\be
\left(g^{\mu\nu}q_\mu q_\nu -m^2\right)=\fr{w^4}{\left(\gb^{\mu\nu}q_\mu q_\nu-\mb^2\right)}~~,
\label{shell3a}
\ee
we see that there is solution for $q_\mu$ that approaches the mass-shell of eq(\ref{shell1}) with 
corrections that are $O(1/q^2)$ where $q$ is a typical component of $q_\mu$~.  
There is an alternative solution in which $q_\mu$ approaches the mass-shell of eq(\ref{shell2})
in a similar manner. At large $q_\mu$ we expect that there is for each mass-shell a solution that lies 
near its positive energy branch and another near the negative energy branch. That is four solutions in all. 
By energy in this context we mean $p^0$ or $\pb^0$ where $p^a=e^{\mu a}q_\mu$ and $\pb^a=\eb^{\mu a}q_\mu$~.
It clear from the "trivial" case in which the lightcones coincide and the theory is
Lorentz invariant that there is a restriction on $w^2$ if all four solutions are to have real
values of $q_0$ for all values of $q_i$~. That is we must have
\be
w^4\le m^2\mb^2~~.
\label{MXB}
\ee
The marginal case 
\be
w^4=m^2\mb^2~~,
\label{constr1}
\ee
results in a massless mode. Breaking the constraint eq(\ref{MXB}) results in solutions with
complex values for $q_0$ for sufficiently small values of $q_i$~.
The same is true in the Lorentz symmetry breaking case.
If we impose condition (\ref{constr1}) then eq(\ref{shell3}) becomes
\be
(g^{\mu\nu}q_\mu q_\nu)(\gb^{\ss\tau}q_\ss q_\tau)-\mb^2g^{\mu\nu} q_\mu q_\nu
                                       -m^2\gb^{\mu\nu}q_\mu q_\nu=0~~.
\label{shell3b}
\ee  
This equation clearly has a solution with $q_\mu=0$~. We can expect to find solutions 
to eq(\ref{shell3b}) of the form $q_\mu=\ll Q_\mu$ that for small $\ll$, satisfy
\be
(\mb^2g^{\mu\nu}+m^2\gb^{\mu\nu})q_\mu q_\nu=0~~.
\ee
That is $q_\mu$ lies on a light-like mass-shell
\be
{\hat g}^{\mu\nu}(u)q_\mu q_\nu=0~~.
\ee
where $u=\mb^2/(m^2+\mb^2)$~. The interpolating metric $\ght^{\mu\nu}(u)$
satisfies $\ght^{\mu\nu}(0)=\gb^{\mu\nu}$ and $\ght^{\mu\nu}(1)=g^{\mu\nu}$~.
It is essential for the causal structure of the theory that ${\hat g}^{\mu\nu}(u)$
retain a proper lightcone structure of the form
\be
{\hat g}^{\mu\nu}(u)=\eta^{ab}{\hat e}^\mu_{~~a}{\hat e}^\nu_{~~b}~~,
\ee
where ${\hat e}^\mu_{~~a}$ is a {\it real} vierbein. In particular we require that
$\det{\hat g}(u)<0$~. Were the determinant to vanish and change sign for $0<u<1$ then no
such vierbein could exist and there would be a breakdown of causality in the theory
at least for values of $m$ and $\mb$ giving rise to a value of $u$ for which $\det {\hat g}^{\mu\nu}(u)>0$.

As we will see in section \ref{quartint} this interpolating cotangent metric is of crucial omportance 
in other interacting theories where similar issues of causality breakdown arise.   

\subsection{\label{MIXWAVE} Wavefunctions for the Mixing Model}

If we have two wave function solutions $F(x)$ and $F'(x)$ of the
equations of of motion eq(\ref{eqmot3}) with
\be
F(x)=\left(\begin{array}{c}
     f(x)\\
     \fb(x)
      \end{array}\right)~~,
\ee
and similarly for $F'(x)$ then the current $J^\mu(x)$ given by (see eq(\ref{current1})
and eq(\ref{current2}))
\be
J^\mu(x)=J_\phi^\mu(x)+J_\psi^\mu(x)~~,
\label{current3}        
\ee 
is conserved even though the individual contributions are not. The mixing term in the
Lagrnagian leads to a flow of probability between the two parts
of the wave function. We can then define a scalar product
\be
(F',F)=\int d^3\bx J^0(x)~~,
\ee
that is independent of time. 

If we choose two plane wave solutions as in eq(\ref{planemix})
we find for the current
\be
J^\mu(x)=\left(\Om A_{q'}^\st A_qg^{\mu\nu}
      +\Omb\Abr_{q'}^\st \Abr_q\gb^{\mu\nu}\right)(q_\nu+q'_\nu)e^{-i(q_\tau-q'_\tau)x^\tau}~~.
\label{current4}
\ee
It is easy to show from eq(\ref{eqmot3}) that, as expected,
\be
\d_\mu J^\mu(x)=-i(q_\mu-q'_\mu)J^\mu(x)=0~~.
\label{current5}
\ee
The scalar product of the two wavefunctions can be obtained from eq(\ref{current4}), thus
\be
(F_{q'},F_q)=(2\pi)^3\dd(\bq-\bq')\left(\Om A_{q'}^\st A_qg^{0\nu}
      +\Omb\Abr_{q'}^\st \Abr_q\gb^{0\nu}\right)(q_\nu+q'_\nu)e^{-i(q_0-q'_0)x^0}~~.
\label{scalar1}
\ee
If $q_0$ and $q'_0$ are on the same branch of the solutions of eq(\ref{shell3}) then
$q_0=q'_0$ when $\bq=\bq'$ and 
\be
(F_{q'},F_q)=(2\pi)^3\dd(\bq-\bq')\left(\Om A_{q}^\st A_qg^{0\nu}
      +\Omb\Abr_{q}^\st \Abr_q\gb^{0\nu}\right)2q_\nu~~.
\label{scalar2}
\ee
If  $q_0$ and $q'_0$ are on different branches of the solutions of eq(\ref{shell3}) then
for consistency we must have
\be
(F_{q'},F_q)=0~~.
\ee
This follows from the fact that when $\bq'=\bq$,
\be
\left(\Om A_{q'}^\st A_qg^{0\nu}+\Omb\Abr_{q'}^\st \Abr_q\gb^{0\nu}\right)(q_\nu+q'_\nu)=0~~,
\ee
which can be proved directly from eq(\ref{eqmot3})~. For definiteness we adopt the following
normalization convention. Denote the mass-shell based on $g^{\mu\nu}$ as the $a$-shell
that based on $\gb^{\mu\nu}$ the $b$-shell. Then if $q$ lies on the branch that asymptotes
the $a$-shell we use the normalization
\be
(F_{q'},F_q)=2\Om g^{0\nu}q_\nu(2\pi)^3\dd(\bq-\bq')~~.
\ee
If $q$ lies on the branch that asymptotes the $b$-shell then
\be
(F_{q'},F_q)=2\Omb \gb^{0\nu}q_\nu(2\pi)^3\dd(\bq-\bq')~~.
\ee

\subsection{\label{MIXQUANT} Quantization of the Mixing Model}

The standard quantization procedure can be applied to the mixing model.
The upshot of course is that we can express the quantum fields as a superposition
of wave functions with coefficients that are annihilation and creation
operators thus
\be
\left(\begin{array}{c}
   \phi(x)\\
\psi(x)
\end{array}\right)
= \int \fr{d^3\bq}{2\Om(2\pi)^3g^{0\nu}q_\nu}a(q)F^{(a)}_q(x)+\int \fr{d^3\bq}{2\Omb(2\pi)^3\gb^{0\nu}q_\nu}b(q)F^{(b)}_q(x)
                                 +\hbox{h.c.}~~.
\ee
where the $a$- and $b$-coefficients satisfy the commutation relations in
eq(\ref{comms1}) and eq(\ref{comms2}) and commute with one another. The Hamiltonian
has the form
\be
H=\int \fr{d^3\bq}{(2\pi)^3\Om2g^{0\nu}q_\nu}q_0a^\dag(q)a(q)+
                     \int \fr{d^3\bq}{(2\pi)^3\Om 2\gb^{0\nu}q_\nu}q_0b^\dag(q)b(q)~~,
\label{HAM4}
\ee
where we have left it implicit that $q_0$ takes values on the branch appropriate to
the integrand in which it appears.

The Feynman propagators are
\begin{eqnarray}
G_F^{\phi\phi}(x-x')&=&-i\la 0|T(\phi(x)\phi(x')|0\ra~~\nonumber\\
G_F^{\phi\psi}(x-x')&=&-i\la 0|T(\phi(x)\psi(x')|0\ra~~,\nonumber\\
G_F^{\psi\psi}(x-x')&=&-i\la 0|T(\psi(x)\psi(x')|0\ra~~,
\end{eqnarray}
and $G_F^{\psi\phi}(x-x')=G_F^{\phi\psi}(x-x')$~. Expressed in terms of Fourier modes
they are
\begin{eqnarray}
G_F^{\phi\phi}(x-x')&=&\fr{1}{\Om}\int\fr{d^4q}{(2\pi)^4}
           \fr{(\gb^{\mu\nu}q_\mu q_\nu-\mb^2)e^{-iq_\tau(x^\tau-x'^\tau)}}{D(q)}\nonumber\\
G_F^{\phi\psi}(x-x')&=&\fr{w^2}{(\Om\Omb)^{1/2}}\int\fr{d^4q}{(2\pi)^4}\fr{e^{-iq_\tau(x^\tau-x'^\tau)}}{D(q)}\nonumber\\
G_F^{\psi\psi}(x-x')&=&\fr{1}{\Omb}\int\fr{d^4q}{(2\pi)^4}
                       \fr{(g^{\mu\nu}q_\mu q_\nu-m^2)e^{-iq_\tau(x^\tau-x'^\tau)}}{D(q)}~~,
\label{PROPAG}
\end{eqnarray}
where
\be
D(q)=(g^{\mu\nu}q_\mu q_\nu-m^2+i\eps)(\gb^{\mu\nu}q_\mu q_\nu-\mb^2+i\eps)-w^4~~.
\ee
As expected each of the three propagators has in Fourier space, poles in
$q_0$ at points where $D(q)$ vanishes. This is of course the determinantal condition
in eq(\ref{shell3})~. To complete the analysis it is necessary to show that these poles
lie on the real $q_0$-axis and that 
$q_0$-integration contour can be threaded through these poles in a way consistent with causality.
The discussion of subsection \ref{MIX} shows that this may not always be possible. In
section \ref{SpecBi} we consider some special cases that illustrate these issues explicitly.

\section{\label{SpecBi}Special Cases of Bi-Metric Backgrounds}

Coleman and Glashow \cite{COLGL1,COLGL2} pointed out that it may be possible to preserve subgroups of 
the Lorentz group even when the full group is no longer an invariance of the theory. These possibilities 
are exemplified by considering the Mixing Interaction model in the following cases.

\subsection{\label{rotinv}Rotational Invariance}

If we choose the vierbeins so that
\be
e^a_{~~\mu}=\left(\begin{array}{cccc}
  \Om&0&0&0\\
  0&1&0&0\\
  0&0&1&0\\
  0&0&0&1\end{array}\right)~~
~~\mbox{and}~~~~
\eb^a_{~~\mu}=\left(\begin{array}{cccc}
  \Omb&0&0&0\\
  0&1&0&0\\
  0&0&1&0\\
  0&0&0&1\end{array}\right)~~,
\ee
then
\be
g_{\mu\nu}=\left(\begin{array}{cccc}
  \Om^2&0&0&0\\
  0&-1&0&0\\
  0&0&-1&0\\
  0&0&0&-1\end{array}\right)~~
~~\mbox{and}~~~~
\gb_{\mu\nu}=\left(\begin{array}{cccc}
  \Omb^2&0&0&0\\
  0&-1&0&0\\
  0&0&-1&0\\
  0&0&0&-1\end{array}\right)~~.
\ee
The two lightcones are then
\begin{eqnarray}
\Om^2(x^0)^2-(x^1)^2-(x^2)^2-(x^3)^2&=&0~~,\nonumber\\
\Omb^2(x^0)^2-(x^1)^2-(x^2)^2-(x^3)^2&=&0~~.
\end{eqnarray}
It is obvious that they share invariance
under the rotation group in the spatial coordinates.

The dispersion relation for the mixing model in this case is
\be
D(q)=
\left(\fr{q_0^2}{\Om^2}-q_1^2-q_2^2-q_3^2-m^2\right)\left(\fr{q_0^2}{\Omb^2}-q_1^2-q_2^2-q_3^2-\mb^2\right)-w^4=0~~.
\label{rotinv1}
\ee
It is obvious that there are two positive real values for $q_0^2$ and hence two positive and two negative solutions
for $q_0$ provided that $w^2\le m\mb$~. Explicitly we have
\be
q_0^2=\fr{1}{2}\left(\Om^2\mu^2+\Omb^2\mub^2\pm\sqrt{(\Om^2\mu^2-\Omb^2\mub^2)^2+4\Om^2\Omb^2w^4}\right)~~,
\ee
where $\mu^2=q_1^2+q_2^2+q_3^2+m^2$ and $\mub^2=q_1^2+q_2^2+q_3^2+\mb^2$~.
In the marginal case where $w^2=m\mb$, the lower branch, for small $q_i$ has the form 
\be
q_0^2=\fr{\Om^2\Omb^2(m^2+\mb^2)}{\Om^2m^2+\Omb^2\mb^2}(q_1^2+q_2^2+q_3^2)~~.
\ee
The implied lightcone structure is perfectly compatible with causality. This is confirmed (rather trivially)
by computing the determinant of the interpolating metric, we have
\be
\det {\hat g}_{\mu\nu}(u)=-(u\Omb^2+(1-u)\Om^2)~~,
\ee
and therefore $\det {\hat g}_{\mu\nu}(u)<0$ for $0<u<1$~.
Finally we note that we can use $D(q)$ in eq(\ref{rotinv1}) in eq(\ref{PROPAG}) to construct the
propagator with Feynman boundary conditions by setting $q_0^2\rightarrow q_0^2+i\eps$~.

\subsection{\label{boostinv} Boost Invariance}

If we choose the vierbeins so that
\be
e^a_{~~\mu}=\left(\begin{array}{cccc}
  1&0&0&0\\
  0&a&0&0\\
  0&0&a&0\\
  0&0&0&1\end{array}\right)~~
~~\mbox{and}~~~~
\eb^a_{~~\mu}=\left(\begin{array}{cccc}
  1&0&0&0\\
  0&\ab&0&0\\
  0&0&\ab&0\\
  0&0&0&1\end{array}\right)~~,
\ee
then $\Om=a^2$ and $\Omb=\ab^2$ and
\be
g_{\mu\nu}=\left(\begin{array}{cccc}
  1&0&0&0\\
  0&-a^2&0&0\\
  0&0&-a^2&0\\
  0&0&0&-1\end{array}\right)~~
~~\mbox{and}~~~~
\gb_{\mu\nu}=\left(\begin{array}{cccc}
  1&0&0&0\\
  0&-\ab^2&0&0\\
  0&0&-\ab^2&0\\
  0&0&0&-1\end{array}\right)~~.
\ee
The two lightcones are then
\begin{eqnarray}
(x^0)^2-a^2(x^1)^2-a^2(x^2)^2-(x^3)^2&=&0~~,\nonumber\\
(x^0)^2-\ab^2(x^1)^2-\ab^2(x^2)^2-(x^3)^2&=&0~~.
\end{eqnarray}
It is obvious that they share invariance
under Lorentz boosts in the 3-direction and rotations in the (1,2)-plane.
Geometrically the two lightcones touch one another in the plane
$x^1=x^2=0$ along the lines $x^0=\pm x^3$~.

The dispersion relation for the mixing model in this case is
\be
\left(q_0^2-\fr{1}{a^2}(q_1^2+q_2^2)-q_3^2-m^2\right)\left(q_0^2-\fr{1}{\ab^2}(q_1^2+q_2^2)-q_3^2-\mb^2\right)-w^4=0~~.
\ee
The solutions for $q_0^2$ are
\begin{eqnarray}
q_0^2&=&\fr{1}{2}\left(\left(\fr{1}{a^2}
                        +\fr{1}{\ab^2}\right)\left(q_1^2+q_2^2\right)+2q_3^2+m^2+\mb^2\right)\nonumber\\
     &&~~~~~~\pm\fr{1}{2}\sqrt{\left(\left(\fr{1}{a^2}
                         -\fr{1}{\ab^2}\right)\left(q_1^2+q_2^2\right)+m^2-\mb^2\right)^2+4w^4}~~.
\end{eqnarray}
It is easily checked that $\det {\hat g}_{\mu\nu}(u)=-(u\ab^2+(1-u)a^2)^2<0$ for $0<u<1$~.
We therefore expect no causality breakdown in this case for any values of $m$ and $\mb$~.

\subsection{\label{sheae} Sheared Lightcones}

If we choose the vierbeins so that
\be
e^a_{~~\mu}=\left(\begin{array}{cccc}
  1&0&0&0\\
  0&1&0&0\\
  0&0&1&0\\
  -\kk&0&0&1\end{array}\right)~~
~~\mbox{and}~~~~
\eb^a_{~~\mu}=\left(\begin{array}{cccc}
  1&0&0&0\\
  0&1&0&0\\
  0&0&1&0\\
  \kk&0&0&1\end{array}\right)~~,
\ee
then $\Om=\Omb=1$ and
\be
g_{\mu\nu}=\left(\begin{array}{cccc}
  1-\kk^2&0&0&\kk\\
  0&-1&0&0\\
  0&0&-1&0\\
  \kk&0&0&-1\end{array}\right)~~
~~\mbox{and}~~~~
\gb_{\mu\nu}=\left(\begin{array}{cccc}
  1-\kk^2&0&0&-\kk\\
  0&-1&0&0\\
  0&0&-1&0\\
  -\kk&0&0&-1\end{array}\right)~~.
\ee
The two lightcones are then
\begin{eqnarray}
(x^0)^2-(x^1)^2-(x^2)^2-(x^3-\kk x^0)^2&=&0~~,\nonumber\\
(x^0)^2-(x^1)^2-(x^2)^2-(x^3+\kk x^0)^2&=&0~~.
\end{eqnarray}
They are obtained from the standard lightcone with $\kk=0$,
by shearing in the 3-direction positively and negatively respectively.

The dispersion relation is
\be
\left((q_0+\kk q_3)^2-q_1^2-q_2^2-q_3^2-m^2\right)\left((q_0-\kk q_3)^2-q_1^2-q_2^2-q_3^2-\mb^2\right)-w^4=0~~.
\ee
Finding an explicit solution for $q_0$ of this quartic equation is difficult in general.
However in the special case for which $\mb=m$ the equation reduces to a quadratic
equation for $q_0^2$ with the result
\be
q_0^2=(\kk^2+1)q_3^2+q_1^2+q_2^2+m^2\pm\sqrt{4\kk^2q_3^2(q_1^2+q_2^2+q_3^2+m^2)+w^4}~~.
\ee
If we consider the marginal case $w=m$ we find for small $q_\mu$ 
\be
q_0^2=q_1^2+q_2^2+(1-\kk^2)q_3^2~~.
\ee
For $\kk<1$ (and $w^2<m^2$) the dispersion relation yields real solutions for $q_0$ for all 
values of $q_1$, $q_2$ and $q_3$ and there are no problems of causality breakdown. However 
when $\kk>1$ we obtain imaginary solutions for $q_0$ and we
no longer have a proper causal structure for the theory. We can confirm this outcome by computing the 
interpolating metric. We have
\be
{\hat g}_{\mu\nu}=\left(\begin{array}{cccc}
  1-\kk^2&0&0&(1-2u)\kk\\
  0&-1&0&0\\
  0&0&-1&0\\
  (1-2u)\kk&0&0&-1\end{array}\right)~~.
\ee
For the equal mass case the relevant value of the interpolating parameter is $u=1/2$~.
We have 
\be
\det{\hat g}_{\mu\nu\\}(u)=-(1-4u(1-u)\kk^2)~~.
\ee
Clearly when $0<\kk<1$ $\det {\hat g}_{\mu\nu}(u)$  remains negative for $0<u<1$~. We expect no causality breakdown
in this case. When $\kk>1$ the determinant will vanish twice in the range $0<u<1$~. When $u$ lies between the 
two zeros we may encounter causality breakdown. The marginal case is $\kk=1$ when a double zero appears at
$u=1/2$~. If we examine the two lightcones we see that for $0<\kk<1$ they do share timelike vectors while
for $\kk>1$ this is no longer the case. In the transitional case with $\kk=1$ the two lightcones touch
along a ray that is the $x^0$-axis. In more complicated theories than the simple mixing model we
will again encounter these issues.

\section{\label{multmix} Mixing model with $N$ Metrics}

We consider briefly a generalisation of the mixing model to the case of $N$ metrics and $N$ fields
and return to the notation of section \ref{MMT}~. We choose as our mixing interaction a contribution to
the Lagrangian of the form
\be
\Lc_I(x)=\fr{1}{2}\sum_{i,j}w_{ij}^2\int d^4x\sqrt{\Om^{(i)}\Om^{(j)}} \phi^{(i)}(x)\phi^{(j)}(x)~~,
\ee
where $w_{ij}=w_{ji}$ and $w_{ii}=0$~. The dispersion relation, a generalisation of eq(\ref{shell3}), is
\be
\det Q(q)=0~~,
\label{shellN}
\ee
where the matrix $Q(q)$ is
\be
Q_{ij}(q)=(g^{(i)\mu\nu}q_\mu q_\nu-m_i^2)\dd_{ij}-w^2_{ij}~~.
\ee
Here $m_i$ is the mass parameter associated with field $\phi^{(i)}(x)$~. The marginal case 
that allows solutions of eq(\ref{shellN}) for arbitrarily small values of $q$ is
\be
\det Q(0)=0~~.
\ee
The dispersion relation for small $q$ then becomes
\be
\sum_iW_ig^{(i)\mu\nu}q_\mu q_\nu=0~~,
\ee
where $W_i$ is the $i^{\mbox{th}}$ diagonal minor of $Q(0)$~. We see that the effective metric
controlling the small $q$ behaviour of the mass shell is the interpolating metric $\ght^{\mu\nu}$
where
\be
\ght^{\mu\nu}=\sum_iu_ig^{(i)\mu\nu}~~,
\ee
with
\be
u_i=\fr{W_i}{W_1+W_2+\cdots}~~.
\ee
We see again that there must be a constraint on the masses such that $\ght^{\mu\nu}$ yields
a proper lightcone. In turn this will impose constraints on the relationships
between the individual lightcones associated with each of the fields. Of course a detailed
analysis of the general case is rather complex. Nevertheless the above discussion shows
the importance of the interpolating metric in understanding the causal structure of the theory.
The interesting question is to what extent these issues re-emerge in a study of more general
models of interacting fields. In the next section we will examine such a model for the case of
two quantum fields.

\section{\label{quartint} Interacting Field Theory - Quartic Interaction}

We can construct a model that includes include scattering effects by allowing quartic interactions for the fields.
A model that is relatively simple but non-trivial has an action of the form
\be
S=S_\phi+S_\psi+S_I~~,
\ee
where
\be
S_\phi=\int d^nx\Om\Lc_\phi(x)~~,
\ee
with
\be
\Lc_\phi(x)=\fr{1}{2}\left(g_0^{\mu\nu}\d_\mu\phi(x)\d_\nu\phi(x)-m_0^2\phi^2(x)\right)~~,
\ee
and
\be
S_\psi=\int d^nx\Omb\Lc_\psi(x)~~.
\ee
with
\be
\Lc_\psi(x)=\fr{1}{2}\left(\gb_0^{\mu\nu}\d_\mu\psi(x)\d_\nu\phi(x)-\mb_0^2\psi^2(x)\right)~~,
\ee
and
\be
S_I=\int d^nx\Lc_I(x)~~,
\ee
with
\be
\Lc_I(x)=
-\fr{\ll_0}{4!}\Om_0(\phi(x))^4-\fr{\llb_0}{4!}\Omb_0(\psi(x))^4 
                            -\fr{\ss_0}{4}(\Om_0\Omb_0)^{\fr{1}{2}}(\phi(x))^2(\psi(x))^2~~.\\
\ee
In the above the squared bare masses for the fields $\phi(x)$ and $\psi(x)$ are, $m_0^2$ and $\mb_0^2$
respectively.  We have also introduced the bare quartic coupling  constants $\ll_0$, $\llb_0$ and $\ss_0$ 
together with bare metrics $g_0^{\mu\nu}$ and $\gb_0^{\mu\nu}$ since we will find that these latter are also 
renormalised. The bare vierbein determinants $\Om_0$ and $\Omb_0$ are related to the bare metrics in an 
obvious way.  Our aim here is to investigate the renormalisation of the bare parameters in
low order of perturbation theory using a dimensional regularization scheme \cite{tHV}. We have therefore 
expressed the action of the theory in terms of $n$-dimensional integrals.

\subsection{\label{FeynRul} Feynman Rules}

We first formulate the perturbation series interms of the bare parameters, expanding
in powers of $\ll_0$ , $\llb_0$ amd $\ss_0$~. The Feynman Rules for computing the terms in the series
for the Green's functions are essentially the same as for the conventional theory
(in which the metrics coincide) except that the vertices and propagators incorporate extra factors
involving powers of $\Om_0$ and $\Omb_0$~.  

In constructing Feynman diagrams we associate the propagation of a $\phi$-field with a solid
line and a factor (see section \ref{SQFT})
$$
iG(q)=\fr{1}{\Om_0}\fr{i}{g_0^{\mu\nu}q_\mu q_\nu-m_0^2+i\eps}~~,
$$
where $q_\mu$ is the momentum flowing through the line. Similarly
the propagation of a $\psi$-field we associate with a dashed line and a factor
$$
i\Gb(q)=\fr{1}{\Omb_0}\fr{i}{\gb_0^{\mu\nu}q_\mu q_\nu-\mb_0^2+i\eps}~~.
$$

The vertices in a Feynman diagram associated with the $\phi^4$, $\psi^4$ and $\phi^2\psi^2$ 
interactions are $-i\ll_0\Om_0$, $-i\llb_0\Omb_0$ and $-i\ss_0(\Om_0\Omb_0)^{1/2}$ respectively.
Of course we enforce momentum conservation at each vertex and finally integrate over loop momentum
$k_\mu$ with a weight
$$
\fr{d^nk}{(2\pi)^n}
$$
Standard symmetry factors are applied as appropriate to each graph.

\subsection{\label{1PI-4} Four Point 1PI-Amplitudes}

To examine the renormalisation properties of the sreies we compute the 1PI-amplitudes derived from
the truncated Green's functions. We will carry out the analysis to second order in the coupling constants.

\begin{figure}[t]
   \centering
  \includegraphics[width=0.6\linewidth]{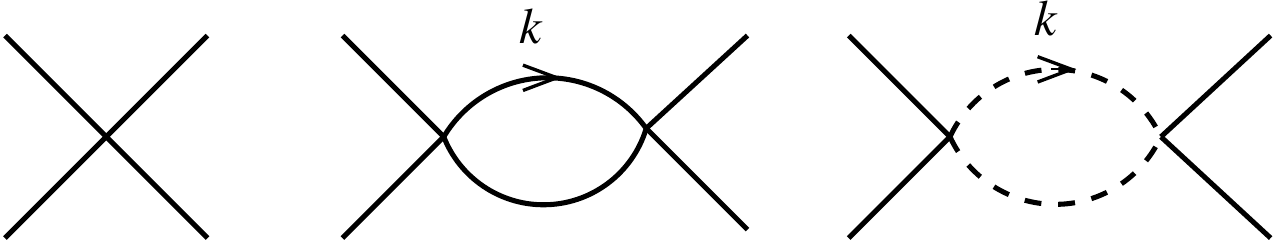}
  \caption{Feynman Diagrams for the amplitude $M_4$}
  \label{FNM}
\end{figure}

The lowest order contributions to the1PI-amplitude for four external $\phi$-lines, $M_4$, are
indicated in Fig \ref{FNM}. Since we are interested only in the divergences of these contributions
it is sufficient to evaluate them with zero momentum on the external lines. Taking into account
the symmetry factors for the loop diagrams and the fact that there are three distinct
ways for attaching the external lines in each case we find
\be
iM_4=-i\ll_0\Om_0+\fr{3}{2}(-i\ll_0)^2I+\fr{3}{2}(-i\ss_0)^2\fr{\Om_0}{\Omb_0}J~~,
\ee
where
\be
I=\int\fr{d^nk}{(2\pi)^n}\left(\fr{i}{g_0^{\mu\nu}k_\mu k_\nu-m_0^2+i\eps}\right)^2
    =\fr{-i\Om_0}{(4\pi)^{n/2}}\GG\left(2-\fr{n}{2}\right)m_0^{(n-4)}~~,
\ee
and
\be
J=\int\fr{d^nk}{(2\pi)^n}\left(\fr{i}{\gb_0^{\mu\nu}k_\mu k_\nu-\mb_0^2+i\eps}\right)^2
    =\fr{-i\Omb_0}{(4\pi)^{n/2}}\GG\left(2-\fr{n}{2}\right)\mb_0^{(n-4)}~~.
\ee

Confining attention to the divergences at $n=4$ this yields
\be
iM_4=-i\Om_0\left[\ll_0+\fr{3}{(4\pi)^2}(\ll_0^2+\ss_0^2)\fr{1}{n-4}\right]~~.
\label{DIV1}
\ee

\begin{figure}[t]
   \centering
  \includegraphics[width=0.7\linewidth]{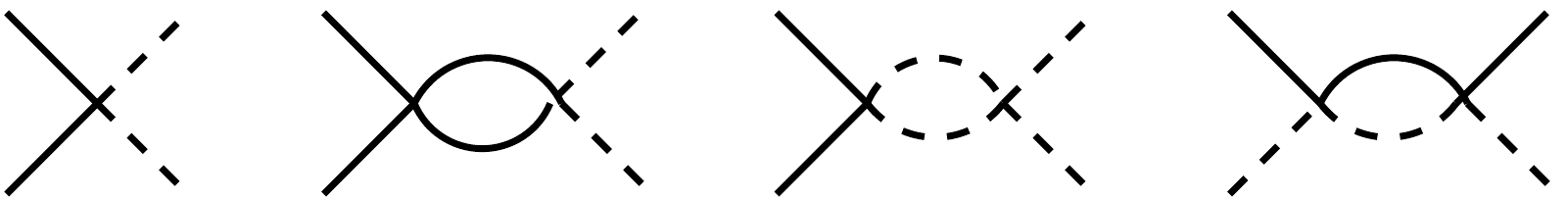}
  \caption{Feynman Diagrams for the amplitude $M_{2,2}$}
  \label{FNM2}
\end{figure}

The corresponding results for the 1PI-amplitude for four external zero momentum $\psi$-lines is $\Mb_4$ where
\be
i\Mb_4=-i\Omb_0\left[\llb_0+\fr{3}{(4\pi)^2}(\llb_0^2+\ss_0^2)\fr{1}{n-4}\right]~~.
\label{DIV2}
\ee
The lowest order contributions to the 1PI-amplitude with two external $\phi$-lines and two
external $\psi$-lines, $M_{2,2}$, are indicated in Fig \ref{FNM2}. Again we set the external lines to zero momentim
we have
\be
\begin{array}{rcl}
iM_{2,2}&=&-i\ss_0(\Om_0\Omb_0)^{1/2}+\fr{1}{2}(-i\ll_0)(-i\ss_0)(\Omb_0/\Om_0)^{1/2}I\\
             &&                 +\fr{1}{2}(-i\llb_0)(-i\ss_0)(\Om_0/\Omb_0)^{1/2}J
                                 +2(-i\ss_0)^2K~~,
\end{array}
\ee
where $I$ and $J$ are defined above and $K$ is given by

\be
K=\int\fr{d^nk}{(2\pi)^n}\fr{i}{g_0^{\mu\nu}k_\mu k_\nu-m_0^2+i\eps}\fr{i}{\gb_0^{\mu\nu}k_\mu k_\nu-\mb_0^2+i\eps}~~.
\ee
We have not given the evaluation of $I$ and $J$ explicitly because it is entirely the same as the corresponding integrals 
in standard theory. However the evaluation of $K$ introduces a new feature involving the interpolating metric
${\hat g}^{\mu\nu}(x)=xg_0^{\mu\nu}+(1-x)\gb_0^{\mu\nu}$~. We therefore do provide an explicit evaluation.
Using the standard representations of the Feynman propagators
\begin{eqnarray}
\fr{i}{g_0^{\mu\nu} k_\mu k_\nu-m_0^2+i\eps}&=&\int_0^{\infty}d\ll_1e^{i\ll_1(g_0^{\mu\nu}k_\mu k_\nu-m_0^2+i\eps)~~,}\nonumber\\
\fr{i}{\gb_0^{\mu\nu} k_\mu k_\nu-\mb_0^2+i\eps}&=&\int_0^{\infty}d\ll_2e^{i\ll_2(\gb_0^{\mu\nu}k_\mu k_\nu-\mb_0^2+i\eps)}~~.
\end{eqnarray}
Making the change of integration variables $\ll_1,\ll_2\rightarrow x,\ll$ where $\ll_1=\ll x$ and $\ll_2=\ll(1-x)$ we find
\be
K=\int_0^1dx\int_0^{\infty}d\ll \ll e^{-i\ll(xm_0^2+(1-x\mb_0^2-i\eps)}
                               \int\fr{d^nk}{(2\pi)^n}e^{i\ll{\hat g}^{\mu\nu}(x)k_\mu k_\nu}~~.
\ee
In order to evaluate the $k$-integral it is necessary to require that ${\hat g}^{\mu\nu}(x)$ does
not become singular for $0<x<1$~. As we have explained in section \ref{INTPL} this requirement is
guaranteed if we demand that the two lightcones do overlap in a manner that permits the existence 
of coordinate rays that are timelike in both metrics. We therfore assume this to be true from now on.
We then obtain 
\be
K=\fr{1}{(2\pi)^n}\int_0^1dx\int_0^{\infty}d\ll\ll e^{-i\ll(xm_0^2+(1-x)\mb_0^2-i\eps)}
                                \fr{i}{\sqrt{-\det{\ght}^{\mu\nu}(x)}}\left(\fr{\pi}{i\ll}\right)^{n/2}~~,
\ee
that is
\be
K=\fr{-i}{(4\pi)^{n/2}}\GG(2-n/2)\int_0^1\fr{dx}{\sqrt{-\det{\ght}^{\mu\nu}(x)}}(xm_0^2+(1-x)\mb_0^2-i\eps)^{n/2-2}~~.
\ee
We are concerned only with the divergence so we may simplify this to
\be
K=\fr{i}{(4\pi)^2}\fr{2}{n-4}\int_0^1\fr{dx}{\sqrt{-\det{\ght}^{\mu\nu}(x)}}~~.
\ee
We have then
\be
iM_{2,2}=-i(\Om_0\Omb_0)^{1/2}\left(\ss_0+\fr{\ss_0\ll_0}{(4\pi)^2}\fr{4}{n-4}
            +\fr{\ss_0^2}{(4\pi)^2}\fr{4\gg}{n-4}\right)~~,
\label{DIV3}
\ee
where 
\be
\gg=\fr{1}{(\Om\Omb)^{1/2}}\int_0^1\fr{dx}{\sqrt{-\det{\ght}^{\mu\nu}(x)}}~~.
\label{ggdef}
\ee

We can now determine the renormalisation structure for the coupling constants. We expand the bare parameters
to second order in renormalised parameters by setting 
\begin{eqnarray}
\ll_0&=&\mu^{4-n}\left(\ll+\fr{a}{n-4}(\ll^2+\ss^2)\right)~~,\nonumber\\
\llb_0&=&\mu^{4-n}\left(\ll+\fr{a}{n-4}(\llb^2+\ss^2)\right)~~,\nonumber\\
\ss_0&=&\mu^{4-n}\left(\ss+\fr{b}{n-4}\ss(\ll+\llb)+\fr{c}{n-4}\ss^2\right)~~,
\label{BCPL1}
\end{eqnarray}
where 
\begin{eqnarray}
a&=&-\fr{3}{(4\pi)^2}~~,\nonumber\\
b&=&-\fr{4}{(4\pi)^2}~~,\nonumber\\
c&=&-\fr{4\gg}{(4\pi)^2}~~.
\label{BCPL2}
\end{eqnarray}
We assume, as will be clear later, that the renormalisation structure for the bare metrics has the form
$g_0^{\mu\nu}=g^{\mu\nu}+O(\ll^2,\ss^2)$  and $\gb_0^{\mu\nu}=\gb^{\mu\nu}+O(\llb^2,\ss^2)$ so
in the above calculation we are free to replace all metric quantities by their renormalised versions.
It then follows that the amplitudes $M_4$, $\Mb_4$ and $M_{2,2}$ are finite to second order in
the renormalised parameters. 

A significant result to emerge from the above analysis is that the the lightcone structure
influences the details of the renormalisation procedure. However if we were to allow the
two metrics to coincide then the above expansion for the bare couplings in terms of the 
renormalized couplings would coicide with the standard result for the theoretical model we are 
investigating. If we allow the two metrics to approach the situation in which their timelike
overlap reduces to zero we would find that the parameter $\gg$ diverges therefore
the renormalisation procedure would exhibit a
singularity in the expansion of the bare parameter $\ss_0$ interms of the renormalised parameters.
The precise physical significance of this is unclear but we obviate the problem by postulating that the 
two metrics must retain a timelike overlap for the quantum field theory to exhibit a proper causal
structure. This is in line with our conclusions from investigating the simple mixing model in 
section \ref{MIX}. 

\subsection{\label{1PI-2} Two Point 1PI-Amplitudes}

We denote the sum over 1PI-amplitudes with two external $\phi$-lines by $\SS(q)$~.
This amplitude contributes to the inverse two point Green's function thus
\be
G_2^{-1}(q)=\Om_0(g_0^{\mu\nu}q_\mu q_\nu-m_0^2)+\SS(q)~~.
\label{INVPROP1}
\ee
Similarly for the 1PI-amplitudes with two external $\psi$-lines we have
\be
\Gb_2^{-1}(q)=\Omb_0(\gb_0^{\mu\nu}q_\mu q_\nu-\mb_0^2)+\SSb(q)~~.
\label{INVPROP2}
\ee

\subsubsection{\label{LOOP1} One Loop Bubble Diagrams}

\begin{figure}[t]
   \centering
  \includegraphics[width=0.4\linewidth]{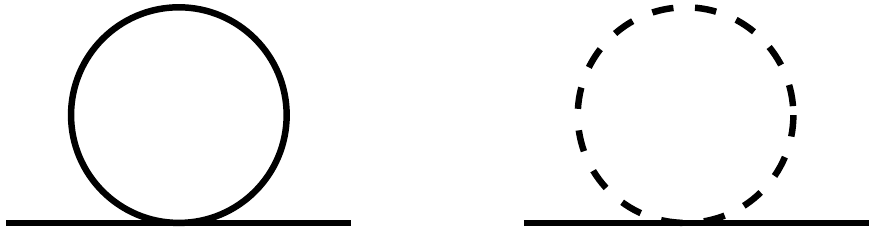}
  \caption{One loop bubble Feynman Diagrams for the amplitude $\SS(q)$}
  \label{FNM3}
\end{figure}

The Feynman diagrams up to one loop, appropriate to computing $i\SS(q)$  are shown in Fig \ref{FNM3}. We have
\be
i\SS(q)= \fr{-i\ll_0}{2}\int \fr{d^nk}{(2\pi)^n}\fr{i}{g_0^{\mu\nu}q_\mu q_\nu-m_0^2+i\eps}
      +\fr{-i\ss_0}{2}\left(\fr{\Om}{\Omb}\right)^{1/2}\int \fr{d^nk}{(2\pi)^n}\fr{i}{g_0^{\mu\nu}q_\mu q_\nu-\mb_0^2+i\eps}~~.
\ee
That is
\be
i\SS(q)=-i\Om_0\fr{\ll_0}{2(4\pi)^{n/2}}\GG(1-n/2)(m_0^2)^{n/2-1}
                   -i(\Om_0\Omb_0)^{1/2}\fr{\ss_0}{2(4\pi)^{n/2}}\GG(1-n/2)(\mb_0^2)^{n/2-1}~~.
\label{SIG1}
\ee

A parallel calculation yields for the 1PI-amplitude for two external $\psi$-lines
\be
i\SSb(q)=-i\Omb_0\fr{\llb_0}{2(4\pi)^{n/2}}\GG(1-n/2)(\mb_0^2)^{n/2-1}
                   -i(\Om_0\Omb_0)^{1/2}\fr{\ss_0}{2(4\pi)^{n/2}}\GG(1-n/2)(m_0^2)^{n/2-1}~~.
\label{SIG2}
\ee

Working to first order in the renormalised masses and coupling constants and retaining only the 
divergent pole terms at $n=4$ we find
\be
i\SS(q)=-i\Om\fr{\ll}{(4\pi)^2}\fr{1}{n-4}m^2-i(\Om\Omb)^{1/2}\fr{\ss}{(4\pi)^2}\fr{1}{n-4}\mb^2~~.
\ee
Here we have introduced the renormalised mass parameters $m^2$ and $\mb^2$ which are
related to the bare masses by the expansion to first order 
\begin{eqnarray}
m_0^2&=&m^2\left(1+\fr{d}{n-4}\ll\right)+\mb^2\left(\fr{f}{n-4}\ss\right)~~,\nonumber\\
\mb_0^2&=&\mb^2\left(1+\fr{\db}{n-4}\llb\right)+m^2\left(\fr{\fb}{n-4}\ss\right)~~.
\end{eqnarray}
We have also replaced $\Om_0$ and $\Omb_0$ by their renormalised values $\Om$ and $\Omb$ since the difference
involves terms of second order in the couplings.
If we choose 
\begin{eqnarray}
d=\db&=&-\fr{1}{(4\pi)^2}~~,\nonumber\\
f&=&-\fr{1}{(4\pi)^2}\left(\fr{\Omb}{\Om}\right)^{1/2}~~,\nonumber\\
\fb&=&-\fr{1}{(4\pi)^2}\left(\fr{\Om}{\Omb}\right)^{1/2}~~,
\end{eqnarray} 
then we see from eq(\ref{INVPROP1}) and eq(\ref{INVPROP2}) that the poles at $n=4$ in the 
expressions for $G_2^{-1}(q)$ and $\Gb_2^{-1}(q)$cancel as required by the dimensional regularization procedure.

For future reference we now use eq(\ref{BCPL1}) and eq(\ref{BCPL2}) to expand the 
right sides of eq(\ref{SIG1}) and eq(\ref{SIG2}) to second order in couplings.
Dropping non-pole terms we list the coefficients of the powers of couplng constants for $i\SS(q)$ as
follows
\begin{eqnarray}
\ll&:&-i\Om\fr{m^2}{(4\pi)^2}\fr{1}{n-4}\nonumber\\
\ss&:&-i(\Om\Omb)^{1/2}\fr{\mb^2}{(4\pi)^2}\fr{1}{n-4}\nonumber\\
\ll^2&:&-i\Om\fr{m^2}{(4\pi)^2}\left(\fr{a+d}{(n-4)^2}+\fr{1}{n-4}\fr{1}{2}
                                                  \left[(a+d)(\log \fr{m^2}{4\pi\mu^2}-\psi(1)-1)+d\right]\right)\nonumber\\
\ss^2&:&-i\Om\fr{m^2}{(4\pi)^2}\left(\fr{a}{(n-4)^2}+\fr{a}{n-4}\fr{1}{2}(\log\fr{m^2}{4\pi\mu^2}-\psi(1)-1)\right)\nonumber\\
&&-i(\Om\Omb)^{1/2}\fr{\mb^2}{(4\pi)^2}\left(\fr{c}{(n-4)^2}+\fr{m^2}{\mb^2}\fr{\fb}{(n-4)^2}\right.\nonumber\\
&&+\left.\fr{1}{n-4}\fr{1}{2}\left[(c+\fr{m^2}{\mb^2}\fb)(\log\fr{\mb^2}{4\pi\mu^2}-\psi(1)-1)+\fr{m^2}{\mb^2}\fb\right]\right)
                                                                                                                  \nonumber\\
\ss\ll&:&-i\Om\fr{m^2}{(4\pi)^2}\left(\fr{\mb^2}{m^2}\fr{f}{(n-4)^2}
                    +\fr{1}{n-4}\fr{1}{2}\left[\fr{\mb^2}{m^2}f(\log\fr{m^2}{4\pi\mu^2}-\psi(1)-1)+\fr{\mb^2}{m^2}f\right]\right)
                                                                                                              \nonumber\\
&& -i(\Om\Omb)^{1/2}\fr{\mb^2}{(4\pi)^2}\left(\fr{b}{(n-4)^2}
                             +\fr{1}{n-4}\fr{1}{2}\left[b(\log\fr{\mb^2}{4\pi\mu^2}-\psi(1)-1)\right]\right)\nonumber\\
\ss\llb&:&-i(\Om\Omb)^{1/2}\fr{\mb^2}{(4\pi)^2}\left(\fr{b+\db}{(n-4)^2}
                      +\fr{1}{n-4}\fr{1}{2}\left[(b+\db)(\log\fr{\mb^2}{4\pi\mu^2}-\psi(1)-1)+\db\right]\right)
\label{SEB1}
\end{eqnarray}
Here $\psi(z)$ is the logarithmic derivative of $\GG(z)$~.

\subsubsection{\label{LOOP2A} Two Loop Bubble Diagrams}

\begin{figure}[t]
   \centering
  \includegraphics[width=0.3\linewidth]{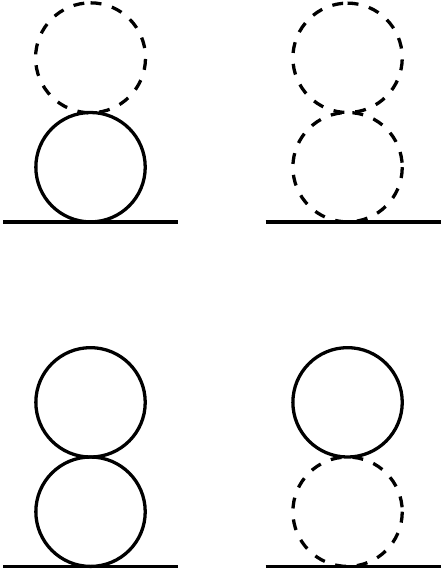}
  \caption{Two loop bubble Feynman Diagrams for the amplitude $\SS(q)$}
  \label{FNM4}
\end{figure}

The two loop bubble diagrams that contribute $i\SS(q)$ are shown in Fig \ref{FNM4}. Because they are 
already second order in the renormalised couplings we can replace all bare quantities by their 
lowest order expansion in renormalised parameters. The loop integrals are easily evaluated
and yield the following results where we list the coefficients of the powers of coupling constants
\begin{eqnarray}
\ll^2&:&i\Om\fr{1}{4}\fr{m^2}{(4\pi)^4}\GG(1-n/2)\GG(2-n/2)\left(\fr{m^2}{4\pi\mu^2}\right)^{n-4}\nonumber\\
\ss^2&:&i\Om\fr{1}{4}\fr{m^2}{(4\pi)^4}\GG(1-n/2)\GG(2-n/2)
                 \left(\fr{m^2}{4\pi\mu^2}\right)^{n/2-2}\left(\fr{\mb^2}{4\pi\mu^2}\right)^{n/2-2}\nonumber\\
\ll\ss&:&i(\Om\Omb)^{1/2}\fr{1}{4}\fr{\mb^2}{(4\pi)^4}\GG(1-n/2)\GG(2-n/2)
                 \left(\fr{m^2}{4\pi\mu^2}\right)^{n/2-2}\left(\fr{\mb^2}{4\pi\mu^2}\right)^{n/2-2}\nonumber\\
\llb\ss&:&i(\Om\Omb)^{1/2}\fr{1}{4}\fr{\mb^2}{(4\pi)^4}\GG(1-n/2)\GG(2-n/2)\left(\fr{\mb^2}{4\pi\mu^2}\right)^{n-4}
\end{eqnarray}
If we drop finite contributions and retain only the poles at $n=4$ we obtain
\begin{eqnarray}
\ll^2&:&-i\Om\fr{m^2}{(4\pi)^4}\left[\fr{1}{(n-4)^2}
                 +\fr{1}{2}\fr{1}{n-4}\left(2\log\fr{m^2}{4\pi\mu^2}-2\psi(1)-1\right)\right]\nonumber\\
\ss^2&:&-i\Om\fr{m^2}{(4\pi)^4}\left[\fr{1}{(n-4)^2}
        +\fr{1}{2}\fr{1}{n-4}\left(\log\fr{m^2}{4\pi\mu^2}+\log\fr{\mb^2}{4\pi\mu^2}-2\psi(1)-1\right)\right]\nonumber\\
\ll\ss&:&-i(\Om\Omb)^{1/2}\fr{\mb^2}{(4\pi)^4}\left[\fr{1}{(n-4)^2}
        +\fr{1}{2}\fr{1}{n-4}\left(\log\fr{m^2}{4\pi\mu^2}+\log\fr{\mb^2}{4\pi\mu^2}-2\psi(1)-1\right)\right]\nonumber\\
\llb\ss&:&-i(\Om\Omb)^{1/2}\fr{\mb^2}{(4\pi)^4}\left[\fr{1}{(n-4)^2}
                 +\fr{1}{2}\fr{1}{n-4}\left(2\log\fr{\mb^2}{4\pi\mu^2}-2\psi(1)-1\right)\right]
\label{SEB2}
\end{eqnarray}

\subsubsection{\label{SUN} Two Loop Sunrise Diagrams}

The two loop sunrise diagrams contributing to $i\SS(q)$ are shown in Fig \ref{FNM5}. Their evaluation
is performed in appendix \ref{RSDS}. The first diagram is $O(\ll^2)$ and gives rise to the pole terms
\begin{eqnarray}
i\SS(q)&=&-i\Om\fr{\ll^2m^2}{(4\pi)^4}\left[\fr{1}{(n-4)^2}+\fr{1}{n-4}(\log\fr{m^2}{4\pi\mu^2}-\psi(1)-3/2)\right]\nonumber\\
   &&      -i\Om\fr{\ll^2}{12(4\pi)^4}\fr{1}{n-4}g^{\mu\nu}q_\mu q_\nu~~.
\label{SUNR1}
\end{eqnarray}
This is the same result for general linear coordinates as in the standard calculation for $\phi^4$-theory. 
The second diagram a similar but more complex outcome. It is convenient to present it in three separate pieces.
The first piece which depends on $\mb^2$, is
\be
i\SS(q)=-i\fr{2\gg\ss^2\mb^2}{(4\pi)^4}\left[\fr{1}{(n-4)^2}
                 +\fr{1}{n-4}\left(\fr{\xi}{\gg}\log\fr{\mb^2}{4\pi\mu^2}-\psi(1)-1\right)\right]~~.
\label{SUNR2}
\ee 
Here the quantity $\xi$ is
\be
\xi=\fr{1}{2(\Om\Omb)^{1/2}}\int_0^1du\fr{\log(1-u)}{\sqrt{-\det\ght^{\mu\nu}(u)}}~~.
\ee

\begin{figure}[t]
   \centering
  \includegraphics[width=0.5\linewidth]{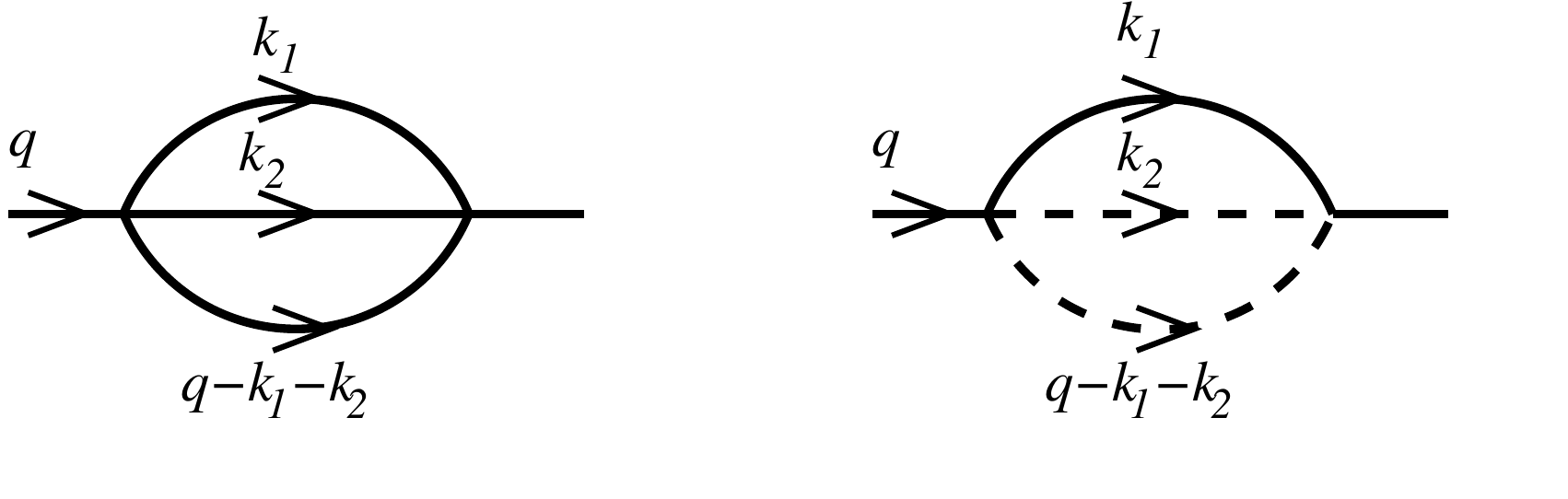}
  \caption{Two loop Rising Sun Feynman Diagrams for the amplitude $\SS(q)$}
  \label{FNM5}
\end{figure}

The second piece which depends on $m^2$ is
\be
i\SS(q)=-i\Om\fr{\ss^2 m^2}{(4\pi)^4}\left[\fr{1}{(n-4)^2}+\fr{1}{n-4}\left(\log\fr{m^2}{4\pi\mu^2}-\psi(1)-2+\zeta\right)\right]~~.
\label{SUNR3}
\ee
where
\be
\zeta=\fr{1}{2\Om}\left[\int_0^1\fr{du}{\sqrt{-\det\ght^{\mu\nu}(u)}}
                                       -\int_0^1\fr{du}{1-u}\left(\fr{1}{\sqrt{-\det\ght^{\mu\nu}(u)}}-\Om\right)\right]~~.
\ee
               
The third term yields the $O(q^2)$-dependent term.
\be
i\SS(q)=-i\Om\fr{\ss^2}{(4\pi)^4}\fr{1}{n-4}\theta^{\mu\nu}q_\mu q_\nu~~,
\label{SUNR4}
\ee
where
\be
\theta^{\mu\nu}=\fr{1}{2\Om}\int_0^1 du\fr{u}{\sqrt{-\det\ght^{\mu\nu}(u)}}(g^{\mu\tau}\ght_{\tau\rho}(u)\gb^{\rho\nu})~~.
\label{SUNR5}
\ee
The matrix $\theta^{\mu\nu}$ is, from its derivation, symmetric. 

\subsubsection{\label{METRICRENORM} Renormalisation of the Metrics}

The contribution to $i\SS(q)$ in eq(\ref{SUNR5}) together with the term $O(q^2)$ in eq(\ref{SUNR1}) 
induce a renormalisation of the bare metric $g_0^{\mu\nu}$~. We set
\be
g_0^{\mu\nu}=g^{\mu\nu}+\fr{1}{n-4}(\ll^2f^{\mu\nu}+\ss^2h^{\mu\nu})~~.
\label{METRN1}
\ee
This implies that
\be
\Om_0=\Om\left(1-\fr{1}{2}\fr{\ll^2g_{\tau\rho}f^{\tau\rho}+\ss^2g_{\tau\rho}h^{\tau\rho}}{n-4}\right)~~.
\label{METRN2}
\ee
Now to $O(q^2)$ we can write, ignoring non-pole contributions,
\be
G_2^{-1}(q)=G_2^{-1}(0)+\Om_0g_0^{\mu\nu}q_\mu q_\nu-\Om\fr{\ll^2}{12(4\pi)^4}\fr{1}{n-4}g^{\mu\nu}q_\mu q_\nu
                         -\Om\fr{\ss^2}{(4\pi)^4}\fr{1}{n-4}\theta^{\mu\nu}q_\mu q_\nu~~.
\ee        
Using eq(\ref{METRN1}) and eq(\ref{METRN2}) we can remove the poles from the $O(q^2)$ term by requiring
\be
f^{\mu\nu}=-\fr{1}{12(4\pi)^4}g^{\mu\nu}~~,
\label{METRN2a}
\ee
and
\be
h^{\mu\nu}=-\fr{1}{2(4\pi)^4}g_{\tau\rho}\theta^{\tau\rho}g^{\mu\nu}+\fr{1}{(4\pi)^4}\theta^{\mu\nu}~~.
\label{METRN2b}
\ee
It is convenient to define $\chi=g_{\mu\nu}\theta^{\mu\nu}$ so that
\be
\chi=\fr{1}{2\Om}\int_0^1\fr{duu}{\sqrt{-\det \ght^{\mu\nu}(u)}}\ght_{\mu\nu}(u)\gb^{\mu\nu}~~.
\ee

Following a parallel calculation we have for the other metric
\be
\gb_0^{\mu\nu}=\gb^{\mu\nu}+\fr{1}{n-4}(\llb^2\fb^{\mu\nu}+\ss^2\hb^{\mu\nu})~~,
\label{METRN3}
\ee
and
\be
\Omb_0=\Omb\left(1-\fr{1}{2}\fr{\llb^2\gb_{\tau\rho}\fb^{\tau\rho}+\ss^2\gb_{\tau\rho}\hb^{\tau\rho}}{n-4}\right)~~.
\label{METRN4}
\ee
We find
\be
\fb^{\mu\nu}=-\fr{1}{12(4\pi)^4}\gb^{\mu\nu}~~,
\label{METRN3a}
\ee
and
\be
\hb^{\mu\nu}=-\fr{1}{2(4\pi)^4}\gb_{\tau\rho}\thb^{\tau\rho}\gb^{\mu\nu}+\fr{1}{(4\pi)^4}\thb^{\mu\nu}~~,
\label{METRN3b}
\ee
where
\be
\thb^{\mu\nu}=\fr{1}{2\Omb}\int_0^1 du\fr{(1-u)}{\sqrt{-\det\ght^{\mu\nu}(u)}}(g^{\mu\tau}\ght_{\tau\rho}(u)\gb^{\rho\nu})~~.
\label{SUNR6}
\ee

It is convenient to define $\chib=\gb_{\mu\nu}\thb^{\mu\nu}$ so that
\be
\chib=\fr{1}{2\Omb}\int_0^1\fr{du(1-u)}{\sqrt{-\det \ght^{\mu\nu}(u)}}\ght_{\mu\nu}(u)g^{\mu\nu}~~.
\ee

\subsubsection{\label{MASSRENORM} Mass Renormalisation}

In order to complete the mass renormalisation to second order in the coupling constants
we first add up all the appropriate contributions to $i\SS(0)$~. Referring back to eq(\ref{SEB1}),
eq(\ref{SEB2}), eq(\ref{SUNR2}), eq(\ref{SUNR3}), and eq(\ref{SUNR4}) we find that the combined result 
for the coefficients of the powers of the renormalised coupling is
\begin{eqnarray}
\ll&:&-i\Om\fr{m^2}{(4\pi)^2}\fr{1}{n-4}~~,\nonumber\\
\ss&:&-i(\Om\Omb)^{1/2}\fr{\mb^2}{(4\pi)^2}\fr{1}{n-4}~~,\nonumber\\
\ll^2&:&-i\Om\fr{m^2}{(4\pi)^4}\left(-\fr{2}{(n-4)^2}-\fr{1}{2}\fr{1}{n-4}\right)~~,\nonumber\\
\ss^2&:&-i\Om\fr{m^2}{(4\pi)^4}\left(\fr{-2}{(n-4)^2}-\fr{1}{n-4}(1-\zeta)\right)\nonumber\\
&&      -i(\Om\Omb)^{1/2}\fr{\mb^2}{(4\pi)^4}\left(\fr{-2\gg}{(n-4)^2}+\fr{2}{n-4}\xi\right)~~,\nonumber\\
\ss\ll&:&-i(\Om\Omb)^{1/2}\fr{\mb^2}{(4\pi)^4}\left(\fr{-1}{(n-4)^2}\right)~~,\nonumber\\
\ss\llb&:&-i(\Om\Omb)^{1/2}\fr{\mb^2}{(4\pi)^4}\left(\fr{-1}{(n-4)^2}\right)~~.
\end{eqnarray}
We then require that $m_0^2$ is a series in the renormalised coupling constants with
coefficients that are poles in $n$ at $n=4$ in such a way that $\Om_0m_0^2-\SS(0,m^2,\mb^2)$
remains finite at $n=4$~. To second order in the couplings this yields
\be
m_0^2=m^2M+\mb^2\left(\fr{\Omb}{\Om}\right)^{1/2}{\bar M}~~,
\ee
where
\begin{eqnarray}
M&=&\left[1-\fr{\ll}{(4\pi)^2}\fr{1}{n-4}+\fr{\ll^2}{(4\pi)^4}\left(\fr{2}{(n-4)^2}
                                                       +\fr{1}{3}\fr{1}{n-4}\right)\right.\nonumber\\
    &&~~~~~~~~~~~~~~~~~~~~~~~~~\left.+\fr{\ss^2}{(4\pi)^4}\left(\fr{2}{(n-4)^2}+\fr{(1-\zeta-\chi/4)}{n-4}\right)\right]~~,
\end{eqnarray}
and
\be
{\bar M}=\left[-\fr{\ss}{(4\pi)^2}\fr{1}{n-4}+\fr{\ss^2}{(4\pi)^4}\left(\fr{2\gg}{(n-4)^2}-\fr{2\xi}{n-4}\right)
                                            +\fr{\ss(\ll+\llb)}{(4\pi)^4}\fr{1}{(n-4)^2}\right]~~.
\ee

The important point is that there is no residual dependence on $\mu$ in the expressions
for the bare masses. It is clear however that as was the case with the coupling constants, 
the renormalisation procedure involves the metrics, through the interpolating metric
as well as directly. 

\subsection{\label{RNGRP} Renormalisation Group}

Strictly speaking we have checked only that the renormalisation procedure works to
second order in the couplings. Although this is a non-trivial result we should, in principle, 
exhibit the result to all orders. This is, as with all field theories, a formidable problem made 
more difficult in our model by the manner in which the metrics enter the process. We shall 
nevertheless assume the correctness of the renormalisation procedure and use it to deduce the 
the renormalisation group equations for the parameters of the theory.

The renormalisation group equations are essentially the statement that the bare parameters
of the theory are invariant under changes of the scale parameter $\mu$~. We have then the
requirement 
\be
\mu\fr{d}{d\mu}\ll_0=\mu\fr{d}{d\mu}\llb_0=\mu\fr{d}{d\mu}\ss_0=0~~.
\ee
In addition in our case we must require 
\be
\mu\fr{d}{d\mu}g_0^{\mu\nu}=\mu\fr{d}{d\mu}\gb_0^{\mu\nu}=0~~.
\ee
If we introduce the variable $t=\log(\mu/\mu_1)$ where $\mu_1$ is a reference scale then from 
eq(\ref{BCPL1}) and eq(\ref{BCPL2}) we find to second order in the coupling constants, the results
\begin{eqnarray}
{\dot\ll}&=&(n-4)\ll+\fr{3}{(4\pi)^2}(\ll^2+\ss^2)~~,\nonumber\\
{\dot\llb}&=&(n-4)\llb+\fr{3}{(4\pi)^2}(\llb^2+\ss^2)~~,\nonumber\\
{\dot\ss}&=&(n-4)\ss+\fr{4}{(4\pi)^2}\ss(\ll+\llb)+\fr{4\gg}{(4\pi)^2}\ss^2~~,
\label{RENG1}
\end{eqnarray}
where $t$-derivatives are indicated by an over dot.
From eq(\ref{METRN1}), eq(\ref{METRN2a}), eq(\ref{METRN2b}), eq(\ref{METRN3}), eq(\ref{METRN3a}) and eq(\ref{METRN3b})
we find again to second order
\begin{eqnarray}
{\dot g^{\tau\rho}}&=&\fr{\ll^2}{6(4\pi)^4}g^{\tau\rho}
                               +\fr{\ss^2}{(4\pi)^4}(\chi g^{\tau\rho}-2\theta^{\tau\rho})~~,\nonumber\\
{\dot\gb^{\tau\rho}}&=&\fr{\llb^2}{6(4\pi)^4}\gb^{\tau\rho}
                                     +\fr{\ss^2}{(4\pi)^4}(\chib\gb^{\tau\rho}-2\thb^{\tau\rho})~~,
\label{RENG2}
\end{eqnarray}
Eqs(\ref{RENG1}) and eqs(\ref{RENG2}) are inter-related because $\gg$ depends on the metrics.

In order to investigate the structure of these equations it is useful to set
$g^{\mu\nu}=A\gg^{\mu\nu}$ and $\gb^{\mu\nu}=B\ggb^{\mu\nu}$ where $\det\gg^{\mu\nu}=\det\ggb^{\mu\nu}=-1$~. 
The equations for the evolution of $A$ and $B$ become
\begin{eqnarray}
{\dot A}=\left(\fr{\ll^2}{6(4\pi)^4}+\fr{\ss^2\chi}{2(4\pi)^4}\right)A~~,\nonumber\\
{\dot B}=\left(\fr{\llb^2}{6(4\pi)^4}+\fr{\ss^2\chib}{2(4\pi)^4}\right)B~~,
\end{eqnarray}
The equations for $\gg^{\mu\nu}$ and $\ggb^{\mu\nu}$ are
\begin{eqnarray}
{\dot\gg}^{\mu\nu}=\fr{\ss^2}{2(4\pi)^4}(\chi\gg^{\mu\nu}-4\Th^{\mu\nu})~~,\nonumber\\
{\dot\ggb}^{\mu\nu}=\fr{\ss^2}{2(4\pi)^4}(\chib\ggb^{\mu\nu}-4\Thb^{\mu\nu})~~,
\label{RGMTRC}
\end{eqnarray}
where $\Th^{\mu\nu}=\th^{\mu\nu}/A$ and $\Thb^{\mu\nu}=\thb^{\mu\nu}/B$~.
It turns out that $\gg$, $\chi$, $\chib$, $\Th^{\mu\nu}$ and $\Thb^{\mu\nu}$ depend
only on $\gg^{\mu\nu}$ and $\ggb^{\mu\nu}$~. From its definition in eq(\ref{ggdef}) we
can show using the change of integration variable
\be
x'=\fr{xA}{xA+(1-x)B}~~,
\ee
that
\be
\gg=\int_0^1\fr{dx'}{\sqrt{-\det\gght^{\mu\nu}(x')}}~~,
\ee
where
\be
\gght^{\mu\nu}(x')=x'\gg^{\mu\nu}+(1-x')\ggb^{\mu\nu}~~.
\ee
Similarly we have
\begin{eqnarray}
\Th^{\mu\nu}&=&\fr{1}{2}\int_0^1\fr{dx'x'}{\sqrt{-\det\gght^{\mu\nu}(x')}}
                                            \gg^{\mu\tau}\gght_{\tau\rho}(x')\ggb^{\rho\nu}~~,\nonumber\\
\Thb^{\mu\nu}&=&\fr{1}{2}\int_0^1\fr{dx'(1-x')}{\sqrt{-\det\gght^{\mu\nu}(x')}}
                                            \gg^{\mu\tau}\gght_{\tau\rho}(x')\ggb^{\rho\nu}~~,
\end{eqnarray}
where $\gght_{\mu\nu}(x')$ is the inverse of $\gght^{\mu\nu}(x')$~. We also have
\begin{eqnarray}
\chi&=&\gg_{\mu\nu}\Th^{\mu\nu}~~\nonumber\\
\chib&=&\ggb_{\mu\nu}\Thb^{\mu\nu}~~.
\end{eqnarray}
We can reduce the renormalisation group equations further by expressing them in a frame
corresponding to the vierbein of one of the metrics. We introduce $E^\mu_{~~a}$
and $\Eb^\mu_{~~a}$ such that
\begin{eqnarray}
\gg^{\mu\nu}&=&E^\mu_{~~a}E^\nu_{~~b}\eta^{ab}~~,\nonumber\\
\ggb^{\mu\nu}&=&\Eb^\mu_{~~a}\Eb^\nu_{~~b}\eta^{ab}~~.
\end{eqnarray}
The two vierbeins are related thus
\begin{eqnarray}
\Eb^\mu_{~~a}&=&E^\mu_{~~b}N^b_{~~a}~~,\nonumber\\
E^\mu_{~~a}&=&\Eb^\mu_{~~b}\Nb^b_{~~a}~~,
\end{eqnarray}
where $\Nb^b_{~~a}$ is the inverse of $N^b_{~~a}$~. It follows that 
\be
N^b_{~~a}=E^b_{~~\mu}\Eb^\mu_{~~a}~~.
\ee
Since $\gg^{\mu\nu}$ is invariant under Lorentz transformations of the vierbein,
that is $E^\mu_{~~a}\rightarrow E^\mu_{~~b}L^b_{~~a}$, we cannot immediately
obtain an equation for ${\dot E}^\mu_{~~a}$ unless we impose a further constraint.
The relevant constraint in this case is to require that
\be
{\dot E}^\mu_{~~a}\eta^{ab}E^\nu_{~~b}=E^\mu_{~~a}\eta^{ab}{\dot E}^\nu_{~~b}~~.
\label{CONSTR}
\ee
The point is of course that were the constraint not to be satisfied it is always possible
to construct a Lorentz transformation $L^a_{~~b}(t)$ that will ensure that eq(\ref{CONSTR})
is satisfied by the transformed vierbein. Similar remarks apply to the construction of 
$\Eb^\mu_{~~a}$~. It follows that
\begin{eqnarray}
{\dot \gg}^{\mu\nu}&=&2{\dot E}^\mu_{~~a}\eta^{ab}E^\nu_{~~b}~~,\nonumber\\
{\dot \ggb}^{\mu\nu}&=&2{\dot\Eb}^\mu_{~~a}\eta^{ab}E^\nu_{~~b}~~,
\end{eqnarray}
It then follows that eq(\ref{RGMTRC}) implies that
\begin{eqnarray}
{\dot E}^\mu_{~~a}&=&\fr{\ss^2}{2(4\pi)^4}(\chi E^\mu_{~~a}-4\Th^{\mu\nu}E^c_{~~\nu}\eta_{ac})~~,\nonumber\\
{\dot\Eb}^\mu_{~~a}&=&\fr{\ss^2}{2(4\pi)^4}(\chib\Eb^\mu_{~~a}-4\Thb^{\mu\nu}\Eb^c_{~~\nu}\eta_{ac})~~,
\label{RGVB}
\end{eqnarray}
We can compute the renormalisation group equation for $N^a_{~~b}$ from eq(\ref{RGVB}) and the result
\be
{\dot N}^b_{~~a}=E^b_{~~\mu}(-{\dot E}^\mu_{~~c}N^c_{~~a}+{\dot\Eb}^\mu_{~~a})~~.
\ee
We have then
\be
{\dot N}^b_{~~a}=\fr{\ss^2}{2(4\pi)^4}((\chib-\chi)N^b_{~~a}+4\Th^{\mu\nu}E^b_{~~\mu}E^d_{~~\nu}\eta_{dc}N^c_{~~a}
                                            -4\Thb^{\mu\nu}E^b_{~~\mu}\Eb^d_{~~\nu}\eta_{da})~~.
\label{RGMTRC1}
\ee
It is convenient to introduce the matrix $\eth^{ab}(u)$ and its inverse $\eth_{ab}(u)$ where
\be
\eth^{ab}(u)=u\eta^{ab}+(1-u)\eta^{cd}N^a_{~~c}N^b_{~~d}~~,
\ee
so that 
\be
\gght^{\mu\nu}(u)=E^\mu_{~~a}E^\nu_{~~b}\eth^{ab}(u)~~.
\ee
It follows that $\det\gght^{\mu\nu}(u)=\det\eth^{ab}(u)$~. The parameter $\gg$ can be expressed as
\be
\gg=\int_0^1\fr{du}{\sqrt{-\det\eth^{ab}(u)}}~~.
\ee
Similarly we find
\begin{eqnarray}
\chi&=&\fr{1}{2}\int_0^1\fr{du u}{\sqrt{-\det\eth^{ab}(u)}}\eta^{ab}N^c_{~~a}N^d_{~~b}\eth_{cd}(u)~~,\nonumber\\
\chib&=&\fr{1}{2}\int_0^1\fr{du(1-u)}{\sqrt{-\det\eth^{ab}(u)}}\eta^{cd}\eth_{cd}(u)~~,
\end{eqnarray}
We find also
\begin{eqnarray}
\Th_{\mu\nu}E^b_{~~\mu}E^d_{~~\nu}&=&\fr{1}{2}\int_0^1\fr{duu}{\sqrt{-\det\eth^{ab}(u)}}
                                       \eta^{bq}\eth_{qp}(u)N^p_{~~r}\eta^{rs}N^d_{~~s}~~,\nonumber\\
\Thb^{\mu\nu}E^b_{~~d}\Eb^d_{~~\nu}\eta_{da}&=&\fr{1}{2}\int_0^1\fr{du(1-u)}{\sqrt{-\det\eth^{ab}(u)}}
                                            \eta^{bq}\eth_{qp}(u)N^p_{~~a}~~.
\label{RGMTRC2}
\end{eqnarray}
From eqs(\ref{RGMTRC1}) and (\ref{RGMTRC2}) we can show that
\be
{\dot N}^b_{~~a}\Nb^a_{~~c}=\fr{\ss^2}{4(4\pi)^4}\int_0^1\fr{du}{\sqrt{-\det\eth^{ab}(u)}}
                                 \eth_{pq}(u)M^{qd}(u)(\dd^p_d\dd^b_c-4\eta^{bp}\eta_{dc})~~,
\label{RGMTRC3}
\ee
where 
\be
M^{qd}(u)=(1-u)\eta^{qd}-uN^q_{~~r}N^d_{~~s}\eta^{rs}~~.
\ee

\subsection{\label{RGFP}Fixed Points of the Renormalisation Group Equations}

It is clear from the results of subsection \ref{RNGRP} that the parameters $\ll$, $\llb$, $\ss$
and $N^a_{~~b}$ satisfy a closed set of differential equations. The evolution of the field
renormalisation variables $A$ and $B$ is determined subsequently. An important aspect
of the renormalisation group equations is the structure of the fixed points.

We can easily check that the right side of eq(\ref{RGMTRC3}) vanishes when $N^b_{~~a}=\dd^b_a$~.
More generally this is true when $N^b_{~~a}$ is a Lorentz transformation. This is to be expected
since the theory then has only the standard lightcone structure and is Lorentz invariant.
The neighbourhood of the this fixed point in $N$-space can be understood by setting $N^b_{~~a}=\dd^b_a+n^b_{~~a}$
where we regard $n^b_{~~a}$ as infinitesimal. From eq((\ref{RGMTRC3}) we find to first order in $n^b_{~~a}$
\be
{\dot n}^b_{~~a}=\fr{2}{3}\fr{\ss^2}{(4\pi)^4}(n^b_{~~a}+n^c_{~~d}\eta^{bd}\eta_{ac})~~.
\label{RGMTRC4a}
\ee
It follows that the difference of the two terms on the right is time independent.
This corresponds to the fact that eq(\ref{RGMTRC3}) is invariant under Lorentz transformations
of $N^b_{~~a}$ from the right. We will simplify the situation and assume that $n^b_{~~a}$
lies in a subspace for which the two terms are equal. We have then
\be
{\dot n}^b_{~~a}=\fr{4}{3}\fr{\ss^2}{(4\pi)^4}n^b_{~~a}~~.
\label{RGMTRC4b}
\ee
The solution is
\be
n^b_{~~a}(t)=n^b_{~~a}(0)\exp\left\{\fr{4}{3}\int_0^tdt'\fr{\ss^2(t')}{(4\pi)^4}\right\}~~.
\label{RGMTRC4c}
\ee
If then $\ss$ were to approach a non-zero fixed point value as $t\rightarrow -\infty$ we could conclude
the theory would acquire Lorentz invariance in the infrared limit. In fact we will see below that 
the situation is somewhat more complicated. 

The analysis of the fixed point structure in the near Lorentz invariant limit is simplified by
noting that to first order in $n^b_{~~a}$ the parameter $\gg=1$ and the renormalisation group equations, 
eq(\ref{RENG1}), for the coupling constants reduce to the standard form for the Lorentz
invariant model. If we adopt the Wilson approach \cite{WILS} and work in $n=4-\eps$ dimensions
then we see that there are a number of fixed points in coupling constant space in the
limit of weak Lorentz symmetry breaking. 

The most obvious fixed point, $A$, lies at the origin of coupling constant space. It is clear from eq(\ref{RENG1}) 
that for $\eps>0$ this fixed point is infrared unstable and correspondingly ultraviolet stable. 
There is a second fixed point, $B$, with $\ss=0$ and 
\be
\ll=\llb=\fr{(4\pi)^2}{3}\eps=2r_*~~,
\ee
where $r_*=((4\pi)^2/6)\eps$.
We can show that this point {\it is} infrared stable and ultraviolet unstable
with respect to all three coupling constants by examining the renormalisation group equations in its 
neighbourhood. Set $\ll=2r_*+u$ and $\llb=2r_*+w$
and treat $u$, $w$ and $\ss$ as small. From eq(\ref{RENG1}) we find
\begin{eqnarray}
{\dot u}&=&\fr{6r_*}{(4\pi)^2}u~~,\nonumber\\
{\dot w}&=&\fr{6r_*}{(4\pi)^2}w~~,\nonumber\\
{\dot \ss}&=&\fr{10r_*}{(4\pi)^2}\ss~~.
\end{eqnarray}
learly $u$, $w$ and $\ss$ all vanish in the infrared limit. In particular
\be
\ss=\ss_1\exp\left\{\fr{10r_*}{(4\pi)^2}t\right\}~~,
\label{RGMTRC5}
\ee
where $\ss_1$ is the value of $\ss$ at $t=0$~. 

\begin{figure}[t]
   \centering
  \includegraphics[width=0.8\linewidth]{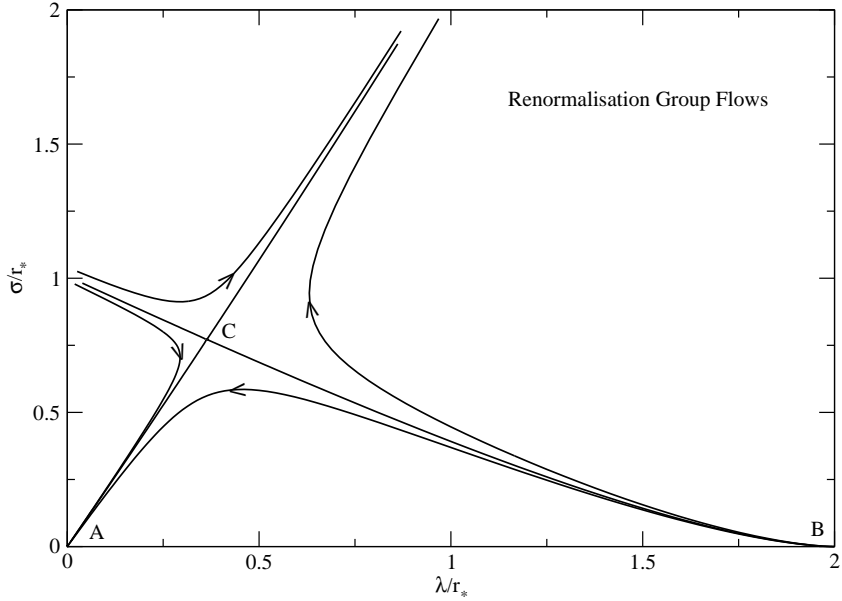}
  \caption{Renormalisation Group flows of the coupling costants, for $\ll=\llb$.
The fixed points are $A$ (ultraviolet stable), $B$ (infrared stable) and
$C$ (ultraviolet stable in the $BC$-direction, infrared stable in the $CA$ direction).}
  \label{RGF}
\end{figure}

There is a third fixed point, $C$, in the domain of positive coupling constants that lies in
the plane $\ll=\llb$ and satisfies the equations
\begin{eqnarray}
\ll^2+\ss^2-2r_*\ll=0~~,\nonumber\\
\fr{4}{3}\ll+\fr{2}{3}\ss-r_*=0~~.
\end{eqnarray}
The solution in the appropriate sector of positive couplings is
\begin{eqnarray}
\ll=\llb=\ll_*&=&0.3611r_*\nonumber\\
\ss=\ss_*&=&0.7718r_*
\end{eqnarray}
By examining the neighbourhood of fixed point $C$ we can easily check that it is ultraviolet stable
in one direction and infrared stable in two other directions. The situation can be largely
understood from Fig \ref{RGF} which exhibits the three fixed points lying in the plane $\ll=\llb$~.
The interesting renormalisation group flows lie along the sides and in the interior of the
"triangle" $ABC$~. The flow along side $BC$ is ultrviolet stable at $C$ the flow along the side
$CA$ is infrared stable. The other flows originate in $B$ and and end in $A$~. The arrow
on each flow indicates the direction of increasing $t$~. If we set the masses $m$ and $\mb$ to zero
for simplicity, we can think of each flow as representing a theory.

A $BA$-flow corresponds to a theory for which the ultraviolet behaviour is controlled by $A$
and the infrared behaviour by $B$~. We can therefore assert that such a theory is aymptotically free
and at large scales will behave as if the effective couplings are $\ll=\llb=2r_*$ and $\ss=0$~.
That is at large distances the two fields $\phi(x)$ and $\psi(x)$ will interact only weakly.
From eq(\ref{RGMTRC4c}) we have
\be
n^b_{~~a}(t)=n^b_{~~a}(0)\exp\left\{\fr{\ss_1^2}{15(4\pi)^2r_*}\exp\left\{\fr{20r_*t}{(4\pi)^2}-1\right\}\right\}~~.
\label{RGMTRC4d}
\ee
In the infrared limit we see that the vanishing of $\ss$ at the fixed point implies
that
\be
n^b_{~~a}(t)=n^b_{~~a}(0)\exp\left\{-\fr{\ss_1^2}{15(4\pi)^2r_*}\right\}~~.
\label{RGMTRC4e}
\ee
It follows that because $\ss$ vanishes in the infrared limit as indicated in eq(\ref{RGMTRC5})
the two lightcones remain distinct.  However to the extent that the two fields $\phi(x)$ and $\psi(x)$ 
are decoupled and each field is invariant under its own version of the Lorentz group. Similar comments apply
in the ultraviolet limit. At intermediate values of $t$ the breakdown of Lorentz invariance shows up becoming maximal when
$\ss$ acquires its largesr value on the trajectory.  

The theory corresponding to the $CA$-flow is ultraviolet free but in the infrared limit is controlled
by the couplings $\ll_*$ and $\ss_*$ corresponding to the fixed point $C$~. Such a theory will exhibit 
a breakdown of Lorentz invariance at large distances. The theory corresponding to the $BC$ flow 
will be controlled in the ultraviolet limit by the parameters of the fixed point $C$ and hence will exhibit 
Lorentz breakdown in this limit. In the infrared limit the theory will again be controlled by the parameter
values at $B$ as discussed above. 

If we pick a theory corresponding to an interior $BA$-flow that starts out from $B$ close to the
$BC$-flow then it will remain close to the $BC$-flow for a large range of $t$ and will spend a
long "time" in the neighbourhood of $C$ thus mimicking the $BC$-theory to high energies
before departing from $C$ and moving close to the $CA$-flow. The corresponding theory therefore
appears to be controlled at high energies by the fixed point $C$ but at very high energies
reconstitutes itself as a $CA$-theory. The intervening stage that is dominated by $C$ exhibits
maximal Lorentz breakdown. Of course the $BA$-flow along the lower edge of the "triangle" $ABC$
corresponds to $\ss=0$ and hence no interaction between the two fields.

When these mass parameters 
are not zero the discussion of the infrared limit will require some modification
to take this into account. We do not deal with this in this paper.
Furthermore when we return to four dimensions, that is $\eps\rightarrow 0$, eq(\ref{RGMTRC4e})
implies that finally $n^b_{~~a}$ vanishes and the two lightcones coincide. In this limit also
the fixed point values of $\ll$ and $\llb$ vanish and the theory becomes trivial as expected
for a quartic scalar theory in four dimensions, as well as Lorentz invariant. However
in the spirit of the $\eps$-expansion \cite{WILS} we might hope that our analysis applies 
at least qualitatively to three-dimensional scalar theories. It would be of great interest
to check this directly by other means.

It is worth noting that the infrared stable fixed point $C$ does not require
$\gg=1$ for its existence. The discussion of the infrared limit in this case
therefore holds good even when the separation of the lightcones is finite.

\subsection{\label{FNTN} Finite Departures from Lorentz Symmetry}

To investigate the possibility of other nontrivial fixed points in general
is not easy. We present two simple cases which can be explored relatively easily.

\subsection{\label{ROTINV} Rotational Invariance}

In the first of these we assume that the matrix $N^a_{~~b}$ relating the two vierbeins
frames in theory has the form
\be
N^a_{~~b}=\left(\begin{array}{cccc}
              \om&0&0&0\\
              0&\om^{-1}&0&0\\
              0&0&\om^{-1}&0\\
              0&0&0&\om
                \end{array}\right)~~.
\ee
This represents a situation where the theory remains invariant under rotations about the 3-axis 
and under boosts along the 3-axis. Eq(\ref{RGMTRC3}) implies
\be
\fr{{\dot\om}}{\om}=-\fr{\ss^2}{(4\pi)^4}\left(\fr{1+\om^4}{1-\om^4}+2\fr{\om^4\log(\om^4)}{(1-\om^4)^2}\right)~~.
\ee
It is easily checked that the coefficient of $\ss^2$ only vanishes when $\om=1$~. This is therefore
only fixed point value for $\om$~. If we set $\om=1+\rho$ and assume $\rho$ is small
then we find
\be
{\dot\rho}=\fr{4}{3}\fr{\ss^2}{(4\pi)^4}\rho~~,
\ee
which is consistent with eq(\ref{RGMTRC4b})~. In this sector of the theory therefore we find
only the fixed point structure already encountered.

\subsection{\label{SHRDLC} Sheared Lightcones}

It is of interest also to exmine a situation in which the relation between the two metrics of the 
theory incorporates a shear of one lightcone relative to the other. In order to achieve this in a
way that is consistent with the renormalisation group equation (\ref{RGMTRC4b}) we set
\be
N^a_{~~b}=\left(\begin{array}{cccc}
               c&0&0&-s/\aa\\
               0&1&0&0\\
               0&0&1&0\\
               -\aa s&0&0&c
               \end{array}\right)~~,
\ee
where $c=\cosh\psi$ and $s=\sinh\psi$ for some hyperbolic angle $\psi$~. The interpolating metric $\eth^{ab}(u)$
is given by
\be
\eth^{ab}(u)=\left(\begin{array}{cccc}
               u+(1-u)(c^2-s^2/\aa^2)&0&0&-(1-u)(\aa-1/\aa)sc\\
                 0&-1&0&0\\
                 0&0&-1&0\\
               -(1-u)(\aa-1/\aa)sc&0&0&u+(1-u)(c^2-\aa^2s^2)\end{array}\right)~~.
\ee
There is no real loss of generality if we assume that $\psi$ is positive and $\aa>1$~.
We have $\det\eth^{ab}=-\DD$ where
\be
\DD=1-u(1-u)(\aa-1/\aa)^2s^2~~.
\ee
This is the form of the determinant for the interpolating metric that arises when the
shear $\kk=(\aa-1/\aa)s$~.
It is obvious that we must impose the constraint $\kk<2$ in order to avoid the
interpolating matrix $\eth^{ab}(u)$ becoming singular for $0<u<1$~. The integrand on the right side 
contains the matrix
eq(\ref{RGMTRC4b})
\be
\eth_{pq}(u)M^{qd}(u)=\left(\begin{array}{cccc}
                         A&0&0&B\\
                         0&(1-2u)&0&0\\
                         0&0&(1-2u)&0\\
                         C&0&0&D\end{array}\right)~~,
\ee
where
\begin{eqnarray}
A&=&\fr{1}{\DD}[-u^2(c^2-s^2/\aa^2)+(1-u)^2(c^2-\aa^2s^2)]~~,\\
B&=&\fr{1}{\DD}(u^2+(1-u)^2)(\aa-1/\aa)sc~~,\\
C&=&-\fr{1}{\DD}(u^2+(1-u)^2)(\aa-1/\aa)sc~~,\\
D&=&\fr{1}{\DD}[-u^2(c^2-\aa^2s^2)+(1-u)^2(c^2-s^2/\aa^2)]~~.
\end{eqnarray}
Using the symmetry $\DD$ under the interchange $u\leftrightarrow (1-u)$ we find
\be
\int_0^1\fr{du}{\sqrt{-\det\eth^{ab}(u)}}\eth_{pq}(u)M^{qd}(u)=\left(\begin{array}{cccc}
                                                          P&0&0&Q\\
                                                          0&0&0&0\\
                                                          0&0&0&0\\
                                                          -Q&0&0&-P\end{array}\right)~~,
\ee
where 
\begin{eqnarray}
P&=&-R(\aa^2-1/\aa^2)s^2~~,\nonumber\\
Q&=&2R(\aa-1/\aa)sc~~,
\end{eqnarray}
with
\be
R=\int_0^1\fr{duu^2}{\DD^{3/2}}~~.
\label{RGMTRC6}
\ee
The renormalisation group equations for $\aa$ and $\psi$ can be inferred from eq(\ref{RGMTRC4b}).
They are
\begin{eqnarray}
\fr{{\dot\aa}}{\aa}&=&\fr{\ss^2}{(4\pi)^4}~~,\nonumber\\ 
\fr{{\dot s}}{s}&=&-\fr{\ss^2}{(4\pi)^4}R(\aa-1/\aa)^2(1+s^2)~~.
\end{eqnarray}
These equations imply a renormalisation group equation for the shear parameter $\kk$ namely
\be
{\dot \kk}=\fr{\ss^2}{(4\pi)^4}R\kk(4-\kk^2)~~.
\label{RGMTRC7}
\ee
It is obvious from eq(\ref{RGMTRC6}) that $R$ depends only on $\kk$ so that 
eq(\ref{RGMTRC7}) involves only the coupling constant $\ss$ and $\kk$~. 
Superficially there appears to be fixed point for $\kk$ at $\kk=2$~.
However it is clear that $R$ has a divergence for this value of $\kk$ and the
leading singularities can be computed as
\be
R\simeq \fr{\pi}{8}\fr{1}{2-\kk}-\fr{1}{4}\log\fr{2}{2-\kk}~~,
\ee
as $\kk\rightarrow 2$ from below. The presence of the pole at $\kk=2$ 
removes the zero in the right side of eq(\ref{RGMTRC6}) and hence any
possibility of a fixed point. Eq(\ref{RGMTRC6}) then implies that in
the infrared limit the shear parameter will renormalise down to zero and 
the theory will again approach the fixed point in which the lightcones
coincide. The presence of the logarithmic singularity at $\kk=2$  means that 
for $\kk>2$ the renormalization group equations of the theory cease to be
real.  A similar divergence is exhibited by the parameter $\gg$	we have
\be
\gg\simeq\fr{1}{2}\log\fr{2}{2-\kk}~~.
\ee
This underlines the point, already evident in the calculation of the perturbation
series for the Green's functions that the the two lightcones must be overlap
in the sense of there being spacetime directions that are timelike in both metrics.
 
The indications from these two examples are that there are no infrared stable fixed points
other than those already discussed. To complete the proof of this
it would be necessary to extend the analysis to the case where $N^a_{~~b}$ 
takes on arbitrary values in the group $SL(4)$ though this is a considerable task 
and beyond the range of this elementary approach.

\section{\label{CONC} Conclusions}

We have investigated simple models of quantum field theory in a multimetric background. 
Our starting point was to seek a causal structure for such theories by requiring that the
various metrics in the model shared a foliation of spacetime that was spacelike with respect
to each of the metrics simultaneously. This suggested the relevance of a parametrisation of the
relationship between metrics that involved a rescaling of the coordinates together with
an appropriate shearing transformation that preserved the spacelike foliation. 

In the case of a bi-metric theory with metrics $g^{\mu\nu}$ and $\gb^{\mu\nu}$ and two 
scalar fields without interaction 
our initial restriction on the metrics was sufficient to permit the combining
of the two quantum fields into a single quantum system. Formally invariance under
Lorentz transformations is lost although the evolution each field is uninfluenced by this.
When a simple transition interaction is introduced the breakdown of Lorentz invariance
is real since each field becomes aware of the other. Although formally the theory
can still be formulated as a causal combined quantum system in fact there are circumstances
that can bring about a breakdown of causality. In assessing the presence of acausality 
we are led to construct the interpolating metric, $\ght^{\mu\nu}(u)=ug^{\mu\nu}+(1-u)\gb_{\mu\nu}$,
with $0<u<1$~. For certain choices of interaction strength this metric, for particular values of
$u$, controls the small momentum behaviour of the dispersion relation. Even though the individual
metrics give rise to conventional light cones the interpolating metric may become singular for
$u$ in its standard range. There will under these circumstances be a range of $u$ for which 
the interpolating metric corresponds to a pseudo-Lorentz metric with diagonal entries
$(+1,+1,-1,-1)$~. This is incompatible with a causally propagating system. In order to
avoid this possibility it is necessary to further restrict the relationship of the two
metrics so that the interpolating metric is never singular for $u$ in its standard range. 
We show that the relevant restriction is that the lightcones of the two metrics overlap
and that therefore there are directions in spacetime that are timelike in both metrics.
When the model is generalised to encompass more than two metrics the same analysis reveals 
the relevance of an interpolating metric that is a positive superposition of all the metrics in the model. 
We have not investigated this situation in detail but the general point remains that to avoid
potential acausality we should require the configuration of metrics to be such that the
interpolating metric remains non-singular.

A bi-metric model in which the fields have quartic interactions is studied. We analyse the
perturbation series for the model to second order in the coupling constants. Dimensional regularisation
is used to deal with the renormalisation of the theory. Two issues stand out. First,
the interpolating metric $\ght^{\mu\nu}(u)$, again plays a crucial role. The evaluation of the Feynman diagrams 
is unambiguous only if $\ght^{\mu\nu}(u)$ remains non-singular throughout the standard range for $u$~. It follows 
that we again must impose the requirement of overlapping lightcones in order to have a causal theory.
Second, given this restriction, it is possible to carry through the renormalisation program, 
at least to second order. However this involves not only the standard renormalisations of
couplings, masses and fields but also the two metrics themselves. We therefore must begin
by formulating the theory with appropriate bare metrics and obtain, scale
dependent, renormalised metrics as part of the calculation.

The renormalisation procedure gives rise to renormalisation group equations for the 
renormalised quantities. The equations for the renormalised couplings and masses are
similar to those that would obtain were the metrics to coincide. They are however
influenced by the relationship of the two metrics. In turn the renormalisation group 
equations for the metrics are of course dependent on the renormalised couplings. We can
separate off the dependence of the overall scale of the metrics and identify it with 
what is conventionally viewed as field renormalisations. We then show that we are free to view
the theory from the vierbein basis of the first metric which then takes the form $\eta^{ab}$ 
while the second metric takes the form $\eta=N^a_{~~c}N^b_{~~d}\eta^{cd}$~. The renormalisation group
equations then condense a little and involve only the couplings the masses, and $N^a_{~~b}$~. 
The matrix $N^a_{~~b}$ involves the scaling and shearing transformations discussed in section \ref{BMG}~.

When $N^a_{~~b}$ is near the identity the renormalisation group equations for the couplings reduce to
those for the Lorentz invariant case. Following Wilson and Kogut \cite{WILS} we work in $4-\eps$ dimensions
and are able to pick out significant fixed points of the these equations that allow us to identify
theories that are well controlled at scales in the ultraviolet and infrared limits. Of course the
well known triviality of quartic scalar theories re-emerges when $\eps\rightarrow 0$ and 
all fixed points return to the origin. However in the spirit of the $\eps$-expansion \cite{WILS}
we might argue that the set of theories is an approximation to a corresponding set of theories
in three dimensions obtained by setting $\eps=1$~. We examine two situations in which $N^a_{~~b}$
departs strongly from the identity. One involves scaling and the other a shearing. In neither case
does it seem likely that there are further fixed points with large Lorentz symmetry breakdown.
The analysis is however not complete. 

It is worth remarking that if we generalise the quartic model in an appropriate way
to involve more metrics and fields we would expect to reproduce the same pattern of restrictions
in the lightcones of the metrics by requiring any two light cones to overlap and share in the manner explained
a set of timelike directions. This is a pairwise condition and does not mean that all metrics
share a timelike directions. However the result emerges from second order perturbation theory 
calculations and may be modified at higher order. Intuition suggests that it is likely
that we will find ultimately that all participating metrics must share in a mutual overlap 
of spacetime vectors that are timelike in each of them. In these circumstances we would expect that
notwithstanding the breakdown of Lorentz invariance there are time variables that permit the
introduction of a Wick rotation and a Euclidean formulation of the theory.

It would be interesting to test these ideas in higher order and in more physical theories
such as the Lorentz non-invariant extensions of QED and the Standard Model \cite{KOST1,KOST2,KOST3,KOST4}
where the interest lies in the energy scale for the onset of Lorentz symmetry breakdown and the 
behaviour of the theory above and below this scale.
 
\section*{Acknowledgements}

I would like to thank Ron Horgan for an illuminating discussion of 
the interpretation of Renormalisation Group flows and Hugh Osborn and Graham Shore
for helpful suggestions and comments.

\appendix

\section{\label{SIM} Singularities of Interpolating Metrics}

Although not completely general it is sufficient for the purposes of demonstrating the result
to make the choice for the two metrics as folows
\be
g_{\mu\nu}=M^\ss_{~~\mu}M^\tau_{~~\nu}\eta_{\ss\tau}~~,
\ee
where
\be
M^\ss_{~~\mu}=\left(
\begin{array}{cccc}
1&0&0&0\\
0&&&\\
0&&\Mc&\\
0&&&
\end{array}
\right)~~,
\ee
where $\Mc$ is a general real $3\times 3$ matrix.
We have then
\be
g_{\mu\nu}=\left(
\begin{array}{cccc}
1&0&0&0\\
0&&&\\
0&&-\Mc^T\Mc&\\
0&&&
\end{array}
\right)~~.
\ee
It is convenient to choose a coordinate basis that diagonalises $\Mc^T\Mc$ leaving
$g_{\mu\nu}$ in the form
\be
g_{\mu\nu}=\left(
\begin{array}{cccc}
1&0&0&0\\
0&-1/a^2&0&0\\
0&0&-1/b^2&0\\
0&0&0&-1/c^2
\end{array}
\right)~~.
\ee
The associated lightcone is
\be
(x^0)^2-\left(\fr{x^1}{a}\right)^2-\left(\fr{x^2}{b}\right)^2-\left(\fr{x^3}{c}\right)^2=0~~.
\ee
The other metric is chosen to be
\be
\gb_{\mu\nu}=S^\ss_{~~\mu}S^\tau_{~~\nu}\eta_{\ss\tau}~~,
\ee
where $S^\ss_{~~\mu}$ is a shear transformation of the form
\be
S^\ss_{~~\mu}=\left(
\begin{array}{cccc}
1&0&0&0\\
-v_1&1&0&0\\
-v_2&0&1&0\\
-v_3&0&0&1
\end{array}
\right)~~.
\ee
We have then
\be
\gb_{\mu\nu}=\left(
\begin{array}{cccc}
1-\bv^2&v_1&v_2&v_3\\
v_1&-1&0&0\\
v_2&0&-1&0\\
v_3&0&0&-1
\end{array}
\right)~~.
\ee
The associated lightcone is
\be
(x^0)^2-(x-v_1x^0)^2-(x^2-v_2x^0)^2-(x^3-v_3x^0)^2=0~~.
\ee
The intersections of the two lightcones with the plane $x^0=1$ are respectively
the ellipsoid centred at the origin
\be
\left(\fr{x^1}{a}\right)^2+\left(\fr{x^2}{b}\right)^2+\left(\fr{x^3}{c}\right)^2=1~~,
\label{ell1}
\ee
and the sphere centred at the point $(x^1,x^2,x^3)=(v_1,v_2,v_3)$
\be
(x-v_1)^2+(x^2-v_2)^2+(x^3-v_3)^2=1~~.
\label{sph1}
\ee

When the centre of the sphere is sufficiently near the origin the sphere and the ellipsoid will overlap.
The points in the interior of the overlap correspond to coordinate rays that are timelike with 
respect to both metrics. When displacement of the sphere is sufficiently large there will be no overlap
with the ellipsoid and the two metrics will not share any timelike rays. The marginal case which is of
interest here is one in which the sphere lies outside the ellipsoid but remains touching it. At the
point of contact the coordinates satisfy both eq(\ref{ell1}) and eq(\ref{sph1}). There is a further
condition that the normals to the surfaces are anti-parallel, that is
\be
\left(
\begin{array}{c}
x^1-v_1\\x^2-v_2\\x^3-v_3
\end{array}\right)=-\ll\left(
\begin{array}{c}
x^1/a^2\\x^2/b^2\\x^3/c^2
\end{array}\right)~~,
\label{norm}
\ee
for some $\ll>0$~.
We can use eq(\ref{norm}) to eliminate the coordinates from eq(\ref{ell1}) and eq(\ref{sph1})
with the results
\be
\fr{1}{a^2}\fr{v_1^2}{(1+\ll/a^2)^2}+\fr{1}{b^2}\fr{v_2^2}{(1+\ll/b^2)^2}+\fr{1}{c^2}\fr{v_3^2}{(1+\ll/c^2)^2}=1~~,
\label{ell2}
\ee
and
\be
\fr{\ll^2}{a^4}\fr{v_1^2}{(1+\ll/a^2)^2}+\fr{\ll^2}{b^4}\fr{v_2^2}{(1+\ll/b^2)^2}+\fr{\ll^2}{c^4}\fr{v_3^2}{(1+\ll/c^2)^2}=1~~.
\label{sph2}
\ee
An appropriate combination of eq(\ref{ell2}) and eq(\ref{sph2}) yields
\be
\fr{1}{a^2}\fr{v_1^2}{(1+\ll/a^2)}+\fr{1}{b^2}\fr{v_2^2}{(1+\ll/b^2)}+\fr{1}{c^2}\fr{v_3^2}{(1+\ll/c^2)}=\fr{1+\ll}{\ll}~~.
\label{comb}
\ee
These equations determine $\ll$ and impose a constraint on $(v_1,v_2,v_3)$~.
The interpolating metric is ${\hat g}_{\mu\nu}(u)=u\gb_{\mu\nu}+(1-u)g_{\mu\nu}$, 
that is
\be
{\hat g}_{\mu\nu}(u)=\left(
\begin{array}{cccc}
1-u\bv^2&uv_1&uv_2&uv_3\\
uv_1&-(u+(1-u)/a^2)&0&0\\
uv_2&0&-(u+(1-u)/b^2)&0\\
uv_3&0&0&-(u+(1-u)/c^2)
\end{array}\right)~~.
\ee
It follows that
\be
\det {\hat g}_{\mu\nu}=-\left(u+\fr{1-u}{a^2}\right)\left(u+\fr{1-u}{b^2}\right)\left(u+\fr{1-u}{c^2}\right)F~~,
\ee
where
\be
F=1-u(1-u)\left[\fr{1}{a^2}\fr{v_1^2}{u+(1-u)/a^2}+\fr{1}{b^2}\fr{v_1^2}{u+(1-u)/b^2}
                                             +\fr{1}{c^2}\fr{v_1^2}{u+(1-u)/c^2}\right]~~.
\ee
The interpolating metric becomes singular when $F$ vanishes. That is when
\be
\fr{1}{a^2}\fr{v_1^2}{1+(1-u)/(ua^2)}+\fr{1}{b^2}\fr{v_1^2}{1+(1-u)/(ub^2)}+\fr{1}{c^2}\fr{v_1^2}{1+(1-u)/(uc^2)}=\fr{1}{1-u}~~.
\label{sing1}
\ee
If we make the identification
\be
\ll=\fr{1-u}{u}~~,
\label{ident}
\ee
then eq(\ref{sing1}) becomes identical to eq(\ref{comb}). The marginal situation that we are investigating
is that $F$ has coincident zeros in $u$~. We ensure this by demanding in addition that
\be
\fr{\d F}{\d u}=0~~.
\ee
This immediately yields eq(\ref{sph2}) after imposing eq(\ref{ident}). Finally we note that eq(\ref{ident}) 
implies that the range  $\ll>0$ corresponds to $0<u<1$ and therefore the requirement of antiparallel normals 
corresponds to the appropriate range for $u$~.

\section{\label{RSDS} Rising Sun Diagrams}

We exhibit the calculation for both of the Rising Sun diagrams in Fig \ref{FNM5}. Because these diagrams are of 
second order in the coupling constants we can replace bare quantities by their lowest order renormalised expressions. 

\subsubsection{\label{RS1} Three Internal $\phi$-lines}

It is useful to discuss the calculation of the first diagram in a certain amount of detail. It 
provides guidance in how to deal with the second less conventional diagram.
Using the Feynman rules we find that the first diagram yields a contribution to $i\SS(q)$ given by
\be
i\SS(q)=-\fr{1}{6}\ll^2(\mu^2)^{4-n}\fr{1}{\Om}L(q,m^2)~~,
\ee
where
\be
L(q,m^2)=\int_0^\infty\fr{d^nk_1d^nk_2}{(2\pi)^{2n}}
                \fr{i}{D(k_1,m^2)}\fr{i}{D(k_2,m^2)}\fr{i}{D(k_1+k_2-q,m^2)}~~,
\ee
with
\be
D(k,m^2)=g^{\mu\nu}k_\mu k_\nu-m^2+i\eps~~.
\ee
From the Schwinger representation of the propagator,
\be
\fr{i}{D(k_1,m^2)}=\int d\ll_1e^{i\ll_1D(k_1,m^2)}~~~~\mbox{etc}~~,
\ee
we have
\begin{eqnarray}
L(q,m^2)&=&\int\fr{d^nk_1d^nk_2}{(2\pi)^{2n}}\int d\ll_1d\ll_2d\ll_3\nonumber\\
      &&         \exp\{i[\ll_1D(k_1,m^2)+\ll_2D(k_2,m^2)+\ll_3D(k_1+k_2-q,m^2)]\}~~.
\end{eqnarray}
We introduce the change of variables $\ll_1=\ll x_1$, $\ll_2=\ll x_2$ and $\ll_3=\ll x_3$ where
$x_1+x_2+x_3=1$~. We have then
\be
L(q,m^2)=\int\fr{d^nk_1d^nk_2}{(2\pi)^{2n}}\int dx_1dx_2dx_3\dd(1-x_1-x_2-x_3)
                  \int d\ll\ll^2e^{i\ll E}~~,
\ee
where 
\be
E=x_1g^{\mu\nu}k_{1\mu}k_{2\nu}+x_2g^{\mu\nu}k_{2\mu}k_{2\nu}+x_3g^{\mu\nu}(k_1+k_2-q)_\mu(k_1+k_2-q)_\nu
                                                                                   -m^2+i\eps~~.
\ee
We displace the origin of the integration variables to the stationary point of $E$ by setting
\begin{eqnarray}
k_1&=&\fr{x_2x_3}{(x_1x_2+x_2x_3+x_3x_1)}q+K_1~~,\nonumber\\
k_2&=&\fr{x_3x_1}{(x_1x_2+x_2x_3+x_3x_1)}q+K_2~~.
\end{eqnarray}
The exponent $E$ takes the form
\be
E=\Ec+\fr{x_1x_2x_3}{(x_1x_2+x_2x_3+x_3x_1)}g^{\mu\nu}q_\mu q_\nu-m^2+i\eps~~,
\ee
where
\be
\Ec=x_1g^{\mu\nu}K_{1\mu}K_{1\nu}+x_2g^{\mu\nu}K_{2\mu}K_{2\nu}+x_3g^{\mu\nu}(K_1+K_2)_\mu(K_1+K_2)_\nu~~.
\ee
The momentum integrations can now be performed
\be
\int\fr{d^nK_1d^nK_2}{(2\pi)^{2n}}e^{i\ll\Ec}=-\fr{\Om^2}{(4\pi)^n}\fr{1}{(i\ll)^n}\fr{1}{(x_1x_2+x_2x_3+x_3x_1)^{n/2}}~~.
\ee
Since our aim is to calculate the ultraviolet poles at $n=4$ we can expand in powers of $q$ 
to second order. We obtain, incorporating the above evaluations,
\be
L(q,m^2)=L(0,m^2)+g^{\mu\nu}q_\mu q_\nu M(m^2)~~,
\ee
where
\be
L(0,m^2)=-i\fr{\Om^2}{(4\pi)^4}\GG(3-n)(m^2)^{n-3}\int \fr{dx_1dx_2dx_3\dd(1-x_1-x_2-x_3)}{(x_1x_2+x_2x_3+x_3x_1)^{n/2}}~~,
\label{RISD1}
\ee
and
\be
M(m^2)=-i\fr{\Om^2}{(4\pi)^4}\GG(4-n)(m^2)^{n-4} 
                  \int\fr{dx_1dx_2dx_3\dd(1-x_1-x_2-x_3)x_1x_2x_3}{(x_1x_2+x_2x_3+x_3x_1)^{n/2+1}}~~.
\label{RISD2}
\ee
As explained by Ramond, the evaluation as it stands, of the integral in eq(\ref{RISD1}) is tricky.
We achieve a more tractable case by first introducing into the integrand the trivial factor $(x_1+x_2+x_3)$,
which evaluates to 1 against the $\dd$-function. We then use the symmetry between $x_1$, $x_2$ and $x_3$
to replace this with $3x_3$. This procedure corresponds to the use of an integration-by-parts identity
introduced by 't Hooft and Veltman. We have then
\be
L(0,m^2)=-i\fr{\Om^2}{(4\pi)^4}\GG(3-n)(m^2)^{n-3}B~~,
\ee
where
\be
B=3\int \fr{dx_1dx_2dx_3\dd(1-x_1-x_2-x_3)x_3}{(x_1x_2+x_2x_3+x_3x_1)^{n/2}}~~,
\label{RISD3}
\ee
To complete the evaluation we eliminate the $\dd$-function by performing the $x_3$-integration,
then make the change of variables $x_1=xy$ and $x_2=(1-x)y$ (the range of all variables is $(0,1)$).
We have 
\be
B=3\int\fr{dxdyy^{1-n/2}(1-y)}{(1-y(1-x(1-x)))^{n/2}}=3\int dxdyy^{1-n/2}+3B'~~,
\ee
where
\be
B'=\int dxdyy^{1-n/2}\left[\fr{1-y}{(1-y(1-x(1-x)))^{n/2}}-1\right]~~.
\ee
For our purposes it is sufficient to evaluate $B'$ at $n=4$~. We find $B'=1$
and hence
\be
B=-\fr{6}{n-4}+3~~,
\ee
and as a result
\be
L(0,m^2)=6i\fr{\Om^2(m^2)^{n-3}}{(4\pi)^4}\fr{1}{(n-4)^2}(1-(n-4)(\psi(1)+3/2))~~.
\ee
From eq(\ref{RISD2}) we see that $M(m^2)$ has only a simple pole at $n=4$~. The residue is contained in
\be
M(m^2)=i\fr{\Om^2(m^2)^{n-4}}{(4\pi)^4}\fr{1}{n-4}A~~,
\ee
where
\be
A=\int\fr{dx_1dx_2dx_3\dd(1-x_1-x_2-x_3)x_1x_2x_3}{(x_1x_2+x_2x_3+x_3x_1)^3}~~.
\ee
Using the change of variables explained above we find
\be
A=\int dxdy\fr{x(1-x)(1-y)}{(1-y(1-x(1-x)))^3}=\fr{1}{2}~~.
\ee
Hence
\be
M(m^2)=i\fr{\Om^2(m^2)^{n-4}}{2(4\pi)^4}\fr{1}{n-4}~~.
\ee
Finally combining these results we find that the pole contributions to $i\SS(q)$ are contained in the result
\be
i\SS(q)=-i\Om\fr{\ll^2m^2}{(4\pi)^4}\left(\fr{m^2}{4\pi\mu^2}\right)^{n-4}\left[1-(n-4)\left(\psi(1)+\fr{3}{2}\right)\right]
                                                                       -i\Om\fr{\ll^2}{12(4\pi)^4}\fr{1}{n-4}g^{\mu\nu}q_\mu q_\nu~~.
\ee
This gives immediately the result in eq(\ref{SUNR1})

\subsubsection{\label{RS2} Two Internal $\psi$-lines and One Internal $\phi$-line}

Applying the Feynman rules to the second diagram in Fig \ref{FNM5} yields a contribution to
$i\SS(q)$ given by
\be
i\SS(q)=-\fr{1}{2}\ss^2(\mu^2)^{4-n}\fr{1}{\Omb}L(q,m^2,\mb^2)~~,
\label{RISD4}
\ee
where
\be
L(q,m^2,\mb^2)=\int\fr{d^nk_1d^nk_2}{(2\pi)^{2n}}\fr{i}{\Db(k_1,\mb^2)}\fr{i}{D(k_2,m^2)}\fr{i}{\Db(k_1+k_2-q,\mb^2)}~~,
\ee
where
\be
\Db(k,\mb^2)=\gb^{\mu\nu}k_\mu k_\nu-\mb^2+i\eps~~.
\ee
Introducing the Schwinger representation for the propagators and making the appropriate changes of integration variables
along the same lines as the previous calculation we obtain
\be
L(q,m^2,\mb^2)=\int\fr{d^nk_1d^nk_2}{(2\pi)^{2n}}\int dx_1dx_2dx_3\dd(1-x_1-x_2-x_3)
                  \int d\ll\ll^2e^{i\ll E}~~,
\ee
where
\begin{eqnarray}
E&=&x_1\gb^{\mu\nu}k_{1\mu}k_{2\nu}+x_2g^{\mu\nu}k_{2\mu}k_{2\nu}\nonumber\\
&&+x_3\gb^{\mu\nu}(k_1+k_2-q)_\mu(k_1+k_2-q)_\nu-(x_1+x_3)\mb^2-x_2m^2+i\eps~~.
\end{eqnarray}
The stationary point of $E$ is given by the requirements
\be
\fr{\d E}{\d k_{1\mu}}=\fr{\d E}{\d k_{2\mu}}=0~~.
\ee
That is
\begin{eqnarray}
x_1\gb^{\mu\nu}k_{1\nu}+x_3\gb^{\mu\nu}(k_1+k_2-q)_\nu&=&0~~,\nonumber\\
x_2g^{\mu\nu}k_{2\nu}+x_3\gb_{\mu\nu}(k_1+k_2-q)_\nu&=&0~~.
\end{eqnarray}
These equations have the solution $k_1=k_1^{(0)}$ and $k_2=k_2^{(0)}$ where
\be
k^{(0)}_{1\mu}=\fr{x_2x_3}{x_1x_2+x_2x_3+x_3x_1}\ght_{\mu\nu}(u)g^{\nu\ss}(u)q_\ss~~,
\ee
and
\be
k^{(0)}_{2\mu}=\fr{x_3x_1}{x_1x_2+x_2x_3+x_3x_1}g_{\mu\ss}\gb^{\ss\tau}\ght_{\tau\rho}(u)g^{\rho\nu}q_\nu~~.
\ee
Note the appearance of the interpolating metric $\ght^{\mu\nu}(u)$~. The argument $u$ is
\be
u=\fr{x_1x_2+x_2x_3}{x_1x_2+x_2x_3+x_3x_1}~~.
\ee
Also
\be
1-u=\fr{x_3x_1}{x_1x_2+x_2x_3+x_3x_1}~~,
\ee
We can then write
\be
k^{(0)}_{2\mu}=(1-u)g_{\mu\ss}\gb^{\ss\tau}\ght_{\tau\rho}(u)g^{\rho\nu}q_\nu~~.
\ee
Using the identity
\be
(ug^{\mu\ss}+(1-u)\gb^{\mu\ss})\ght_{\ss\nu}(u)=\dd^\mu_\nu~~,
\label{IDEN}
\ee
we can show that
\be
k^{(0)}_{2\mu}=q_\mu-u\ght_{\mu\rho}g^{\rho\nu}q_\nu~~.
\ee
We then have
\be
(k^{(0)}_1+k^{(0)}_2-q)_\mu=-\fr{x_1x_2}{x_1x_2+x_2x_3+x_3x_1}\ght_{\mu\rho}g^{\rho\nu}q_\nu~~.
\ee
We shift the origin of the momentum integration variables 
by setting
\begin{eqnarray}
k_{1\mu}=k^{(0)}_{1\mu}+K_{1\mu}~~,\nonumber\\
k_{2\mu}=k^{(0)}_{2\mu}+K_{2\mu}~~.
\end{eqnarray}
We have then
\be
E=\Ec+\Ec_R-(x_1+x_3)\mb^2-x_2m^2+i\eps~~,
\ee
where
\be
\Ec=x_1\gb^{\mu\nu}K_{1\mu}K_{1\nu}+x_2g^{\mu\nu}K_{2\mu}K_{2\nu}+x_3\gb^{\mu\nu}(K_1+K_2)_\mu(K_1+K_2)_\nu~~,
\ee
and the remainder, $\Ec_R$, is of course the value of $E$ at the stationary point and is given by
\be
\Ec_R=x_1\gb^{\mu\nu}k^{(0)}_{1\mu}k^{(0)}_{1\nu}+x_2\gb^{\mu\nu}k^{(0)}_{2\mu}k^{(0)}_{2\nu}
                            +x_3\gb^{\mu\nu}(k^{(0)}_1+k^{(0)}_2-q)_\mu(k^{(0)}_1+k^{(0)}_2-q)_\nu~~.
\ee

Note that $\Ec_R=O(q^2)$ for small $q$~.

After performing the momentum integrations we have
\be
\int\fr{d^nK_1d^nK_2}{(2\pi)^{2n}}e^{i\ll\Ec}=
   -\fr{\Omb}{(4\pi)^n}(i\ll)^{-n}\fr{1}{(x_1x_2+x_2x_3+x_3x_1)^{n/2}}\fr{1}{\sqrt{-\det\ght^{\mu\nu}(u)}}~~.
\ee
Since we only wish to compute the poles at $n=4$ in $L(q,m^2,\mb^2)$ it is sufficient to consider 
the low $q$ expansion
\be
L(q,m^2,\mb^2)=L(0,m^2,\mb^2)+M^{\mu\nu}q_\mu q_\nu~~.
\ee
We obtain these terms by expanding
\be
e^{i\ll\Ec_R}=1+i\ll\Ec_R~~,
\ee
which is correct to $O(q^2)$~. Performing the $\ll$ integration we find
\begin{eqnarray}
L(0,m^2,\mb^2)&=&-i\fr{\Omb}{(4\pi)^n}\GG(3-n)\int \fr{dx_1dx_2dx_3\dd(1-x_1-x_2-x_3)}{\sqrt{-\det\ght^{\mu\nu}(u)}}\nonumber\\
 &&~~~~~~~~~~~~~~~~~~~~~~~~~~~~~~~~~~~~        \fr{[(x_1+x_3)\mb^2+x_2m^2]^{n-3}}{(x_1x_2+x_2x_3+x_3x_1)^{n/2}}~~,
\end{eqnarray}
and
\begin{eqnarray}
M^{\mu\nu}q_\mu q_\nu&=&-i\fr{\Omb}{(4\pi)^n}\GG(4-n)\int i
                                 \fr{dx_1dx_2dx_3\dd(1-x_1-x_2-x_3)}{\sqrt{-\det\ght^{\mu\nu}(u)}}\nonumber\\
 &&~~~~~~~~~~~~~~~~~~~~~~~~~~~~~~~~~~~~        \fr{[(x_1+x_3)\mb^2+x_2m^2]^{n-4}}{(x_1x_2+x_2x_3+x_3x_1)^{n/2}}\Ec_R~~.
\end{eqnarray}
The evaluation of $\Ec_R$ is as follows. Using the results for $k^{(0)}_1$ and $k^{(0)}_2$ we find
\begin{eqnarray}
\Ec_R&=&\fr{x_1x_2x_3}{x_1x_2+x_2x_3+x_3x_1}[u\gb^{\mu\nu}\ght_{\mu\tau}(u)g^{\tau\ss}q_\ss\ght_{\nu\eps}g^{\eps\rho}q_\rho
                                                                                          \nonumber\\
&&~~~~~~~~~~~~~~~~~~~~~~~~~~~~~~~~~~~~
   (1-u)\gb^{\nu\tau}\ght_{\tau\eps}g^{\eps\ss}q_\ss g_{\nu\aa}\gb^{\aa\ll}\ght_{\ll\bb}(u)g^{\bb\rho}q_\rho]~~.
\end{eqnarray}
Using the identity in eq(\ref{IDEN}) this becomes
\be
\Ec_R=\fr{x_1x_2x_3}{x_1x_2+x_2x_3+x_3x_1}(g^{\ss\tau}\ght_{\tau\mu}(u)\gb^{\mu\rho})q_\ss q_\rho~~.
\ee
The matrix $(g^{\ss\tau}\ght_{\tau\mu}(u)\gb^{\mu\rho})$ can be shown explicitly to be symmetric.

It is convenient to re-express $L(0,m^2,\mb^2)$ as a sum of two terms in the form
\be
L(0,m^2,\mb^2)=-i\fr{\Om}{(4\pi)^4}\GG(3-n)[\mb^2 L_1+m^2L_2]~~,
\ee
where
\begin{eqnarray}
L_1&=&\int \fr{dx_1dx_2dx_3\dd(1-x_1-x_2-x_3)}{\sqrt{-\det\ght^{\mu\nu}(u)}}\nonumber\\
 &&~~~~~~~~~~~~~~~~~~~~~~~~~~~~~~~~~~~~        \fr{(x_1+x_3)[(x_1+x_3)\mb^2+x_2m^2]^{n-4}}{(x_1x_2+x_2x_3+x_3x_1)^{n/2}}~~,
\end{eqnarray}
and
\begin{eqnarray}
L_2&=&\int \fr{dx_1dx_2dx_3\dd(1-x_1-x_2-x_3)}{\sqrt{-\det\ght^{\mu\nu}(u)}}\nonumber\\
 &&~~~~~~~~~~~~~~~~~~~~~~~~~~~~~~~~~~~~        \fr{x_2[(x_1+x_3)\mb^2+x_2m^2]^{n-4}}{(x_1x_2+x_2x_3+x_3x_1)^{n/2}}~~,
\end{eqnarray}
We now eliminate the $\dd$-functions by integrating over $x_2$ and introduce the change of variables
$x_1=xy$ and $x_3=(1-x)y$ with the result
\be
u=\fr{1-y}{1-y(1-x(1-x)}~~,
\label{CHV}
\ee
and
\be
L_1=\int\fr{dxdyy^{2-n/2}[y\mb^2+(1-y)m^2]^{n-4}}{\sqrt{-\det\ght^{\mu\nu}(u)}(1-y(1-x(1-x)))^{n/2}}~~.
\ee
We invert eq(\ref{CHV}) to replace $y$ by $u$ as the integration variable and obtain, ignoring
contributions that vanish when $n=4$,
\be
L_1=(\mb^2)^{n-4}\int\fr{dxdux^{1-n/2}(1-x)^{1-n/2}(1-u)^{n/2-2}}{\sqrt{-\det\ght^{\mu\nu}(u)}}~~,
\ee
with the result
\be
L_1=\fr{(\GG(2-n/2))^2}{\GG(4-n)}(\mb^2)^{n-4}\int\fr{du(1-u)^{n/2-2}}{\sqrt{-\det\ght^{\mu\nu}(u)}}~~.
\ee
By a parallel calculation we find
\be
L_2=\fr{(\GG(n/2-1))^2}{\GG(n-2)}(m^2)^{n-4}\int\fr{duu(1-u)^{1-n/2}}{\sqrt{-\det\ght^{\mu\nu}(u)}}~~.
\ee
We have also
\be
M^{\mu\nu}q_\mu q_\nu=i\Omb\fr{1}{n-4}\fr{1}{(4\pi)^4}
                  \int\fr{duu}{\sqrt{-\det\ght^{\mu\nu}(u)}}(g^{\ss\tau}\ght_{\tau\mu}(u)\gb^{\mu\rho})q_\ss q_\rho~~.
\ee
Combining these results with eq(\ref{RISD4}) and retaining only the pole terms at $n=4$ we obtain the results in 
eq(\ref{SUNR2}), eq(\ref{SUNR3}) and eq(\ref{SUNR4}).

\bibliography{mm2}
\bibliographystyle{unsrt}

\end{document}